# Day-to-Day Dynamic Traffic Assignment with Imperfect Information, Bounded Rationality and Information Sharing

Yang Yu[a]     Ke Han[b,a,*]     Washington Ochieng[a]

[a]Department of Civil and Environmental Engineering, Imperial College London, United Kingdom
[b]School of Transportation and Logistics, Southwest Jiaotong Uinversity
*Corresponding author, email: kxh323@gmail.com

## Abstract

This paper presents a doubly dynamic day-to-day (DTD) traffic assignment model with simultaneous route-and-departure-time (SRDT) choices while incorporating incomplete and imperfect information as well as bounded rationality. Two SRDT choice models are proposed to incorporate imperfect travel information: One based on multinomial Logit (MNL) model and the other on sequential, mixed multinomial/nested Logit model. These two variants, serving as base models, are further extended with two features: bounded rationality (BR) and information sharing. BR is considered by incorporating the indifference band into the random utility component of the MNL model, forming a BR-based DTD stochastic model. A macroscopic model of travel information sharing is integrated into the DTD dynamics to account for the impact of incomplete information on travelers' SRDT choices. These DTD choice models are combined with within-day dynamics following the Lighthill-Whitham-Richards (LWR) fluid dynamic network loading model. Simulations on large-scale networks (Anaheim) illustrate the interactions between users' adaptive decision making and network conditions (including local disruption) with different levels of information availability and user behavior. Our findings highlight the need for modeling network transient and disequilibriated states, which are often overlooked in equilibrium-constrained network design and optimization. The MATLAB package and computational examples are available at https://github.com/DrKeHan/DTD

## 1. Introduction

Dynamic traffic assignment (DTA) models aim to describe and predict time-varying traffic flows on networks consistent with established travel demand, travel behavior, and traffic flow theory. A widely accepted classification of DTA models is influenced by Wardrop's principles (Wardrop, 1952), known as the dynamic extensions of system optimal (SO) and user equilibrium (UE). The reader is referred to Peeta and Ziliaskopoulos (2001), Boyce et al. (2001), Szeto and Lo (2005, 2006), and Bliemer et al. (2017) for a comprehensive review of these DTA models.

Another perspective of differentiating DTA models concerns with the time scales involved in the traffic dynamics; namely within-day and day-to-day DTA models. Within-day models are typically associated with a single time horizon within a calendar day, such as morning peak hours, assuming that analyses carried out therein can be transferred to multiple days under the same, unperturbed network conditions (e.g. travel demand and network properties). Day-to-day (DTD) models, on the other hand, are concerned with the evolutionary nature of traffic on a sequence of days, which is influenced by the evolving network properties and travelers' adaptive learning and decision making. Here, the notion of 'day' is broadly interpreted to mean an epoch, be it a week, month or arbitrary period in which traffic undergoes a discernible change.

While dynamic system optimal and dynamic user equilibrium, which fall within the category of within-day models, are by far the most widely studied forms of DTA, there is a strong case for investigating DTD dynamic traffic models under disequilibrium conditions. Indeed, the equilibrium state may not exist in real-world traffic networks since it can be easily disturbed by varying travel demand (such as weather, special events, and departure time flexibility) and constant network perturbations (such as traffic incidents, construction works, adaptive traffic controls), which could lead to travelers' route and departure time uncertainties and result in the daily fluctuations of network flow patterns. Instead of attempting to predict the unperturbed network equilibrium, DTD DTA models aim to describe travelers' learning, adjustment, and decision-making behavior on both within-day and day-to-day time scales. This modeling perspective is crucial for capturing network transient states as a result of abrupt network changes, or fluctuations near an equilibrium given complex interaction of information and decision making. DTD DTA models aim to describe and predict the traffic disequilibrium processes, understanding travelers' learning processes and adaptive behavior, while remain flexible in modeling network disruptions and incorporating various information provision and feedback mechanisms.

Day-to-day traffic assignment models can be categorized into the following:
1. Deterministic processes based on deterministic choice models (Nagurney and Zhang, 1997; Friesz et al., 1994; Yang and Zhang, 2009; He et al., 2010; Smith and Mounce, 2011; Zhou et al., 2017);
2. Deterministic processes based on probabilistic/stochastic choice models (e.g. random utility theory) (Horowitz, 1984; Cantarella and Cascetta, 1995; Watling, 1999); and
3. Stochastic processes based on probabilistic/stochastic choice models (Cascetta, 1989; Cantarella and Cascetta, 1995; Watling and Hazelton, 2003; Watling and Cantarella, 2013).

According to Cantarella and Watling (2016), stochastic models are more naturally associated with modelling the variability that is seen to occur in real-life systems, which are able to represent both dynamic transitions and steady-state fluctuations not seen in equilibrium models. A significant number of studies are associated with DTD traffic network modeling (Guo and Liu, 2011; Cantarella and Watling, 2016; Wang et al., 2016; Xiao and Lo, 2016; Bifulo et al., 2016; Rambha and Boyles, 2016; Zhang et al., 2018; Guo and Szeto, 2018; Watling and Hazelton, 2018).

The modeling of traveler's route and/or departure time choices in the DTA literature often assumes that the travelers have *perfect* and *complete* knowledge of the traffic system and behave in a totally *rational* manner. This means that travelers have access to the actual experienced costs associated with all travel choices and only choose those with minimum costs. Such an assumption has formed the basis of many deterministic DTD dynamics (Smith 1984; Nagurney and Zhang 1997; Friesz et al., 1994; He and Liu, 2012; Bie and Lo 2010; Guo et al. 2015), which have their limitations due to the lack of complete and accurate information on all the alternatives, and individual perception errors of the same situation.

Stochastic/probabilistic choice models are widely studied in DTD assignment to incorporate imperfect information as well as perception heterogeneity based on random utility theory (Cascetta, 1989; Watling, 1996; Hazelton and Walting, 2004; Watling and Cantarella, 2013; Parry and Hazelton, 2013). In contrast, incomplete information and bounded rationality (BR) are less studied and understood in stochastic modeling. Indeed, only a few recent studies aim to incorporate information sharing behavior in day-to-day choice models (Iryo, 2016; Xiao and Lo, 2016; Wei et al., 2016; Shang et al., 2017; Zhang et al., 2018). However, most of these models investigate information sharing from an agent-based (i.e. microscopic) perspective, which are demonstrated on simple networks. Few have proposed generalizable and computationally efficient macroscopic modeling counterpart that is immediately suitable for simulating large-scale dynamic traffic networks.

Bounded rationality (BR) is an important generalization of choice modeling that allows sub-optimal alternatives to be chosen within an indifference band (Mahmassani and Chang, 1987). While a number of studies have incorporated BR into the DTD framework, they all focus on route choice (Guo and Liu, 2011; Di et al., 2015; Ye and Yang, 2017) or departure time choice (Guo et al., 2017) separately; none has considered doubly dynamic model with *simultaneous route-and-departure-time* (SRDT) choices. In addition, and more importantly, these DTD studies with BR all employ a deterministic approach assuming perfect and complete information, which is yet to be generalized in a stochastic context.

Significant effort has been dedicated to developing behaviorally sound (and sometimes sophisticated) DTD choice models with relevant considerations of information availability and user heterogeneity. However, few studies employ realistic within-day traffic dynamics or network examples beyond small-size problems (e.g. a bottleneck). Within-day dynamics, on the supply side of the traffic system, are just as important since they inform and influence travelers' decisions on an iterative basis. Cascetta and Cantarella (1991) employ a traffic queuing model to describe link and network delays. Friesz et al. (1994) and Balijepalli et al. (2007) employ the affine link delay function to propagate link flows and delays. Iryo (2016), Guo et al. (2017) and Liu et al. (2017) focus on a single bottleneck following Vickrey's queuing model. Large-scale implementation of such doubly dynamic models, while capturing important and realistic congestion phenomena, remains a key step towards their applications.

Aiming to address the aforementioned gaps in the literature, this paper proposes a doubly dynamic traffic assignment (DDTA) model with realistic user behavior and network dynamics. The analytical and deterministic model is flexible in incorporating further model extensions such as bounded rationality, and incomplete information. In particular, the following contributions are highlighted (comprehensive literature reviews on individual topics are provided in Section 2).

1. **Doubly dynamic traffic assignment model with SRDT choices (SRDT-DDTA).** We propose a macroscopic DTD model with simultaneous-route-and-departure-time (SRDT) choices, where the within-day dynamics follow the Lighthill-Whitham-Richards (LWR) fluid dynamic network loading model. According to the literature review in Section 2, this is the first doubly dynamic model with SRDT choices. The proposed model allows a realistic representation of travelers' choice set in response to network conditions and changes.

2. **Realistic traffic dynamics and transferability.** We employ the LWR-based DNL procedure for describing the within-day traffic dynamics on large-scale traffic networks (e.g. the Anaheim network) while capturing realistic traffic phenomena such as shock waves and vehicle spillback. This is crucial for analyzing real-world networks with constant supply shortage due to recurrent or incidental disruptions. The Matlab codes are made openly available to help advance the literature of doubly dynamic traffic assignment beyond small-scale and illustrative numerical examples.

3. **Model extension with bounded rationality.** The SRDT-DDTA model is extended with bounded rationality, by incorporating the indifference band in the random perception errors in the DTD model. This leads to the first BR-based DDTA model with SRDT choices.

4. **Model extension with macroscopic information sharing mechanism.** The SRDT-DDTA model is further extended with a model of travel information sharing to account for the effect of incomplete information on travelers' SRDT choices. This is done at the macroscopic level for computational efficiency, and consistent with the SRDT choices and fluid-based dynamic network loading. This is suitable for large-scale simulations with generalizable insights not easily accessible from agent-based simulations.

We conduct a battery of sensitivity and scenario-based analyses on the Sioux Falls network (530 O-D pairs, 6,180 paths, 30,000 trips) and Anaheim network (1,406 O-D pairs, 30,719 paths, 30,000 trips). Local capacity disruption and restoration are simulated to highlight the need for explicitly modeling SRDT choices in DTD dynamics, and understand the interaction between travelers' decision making and traffic dynamics with different levels of information availability and user behavior; see Sections 5 and 6 for detailed discussion and insights. The proposed doubly dynamics contribute to state of the art by presenting a unified framework for modeling realistic user and traffic dynamics under transient or non-equilibrium states, which offers useful tools for network management and policy appraisal.

We note that the primary goal of this paper is to propose a doubly dynamic model for general and large-scale networks where travelers' choices include departure time and route (Contributions 1 & 2 above). The bounded rationality and information sharing (Contributions 3 & 4) are examples of extensions that can be easily made, and demonstrate the range of behavioral realism that the model can accommodate. Our intent is not to formulate overly complicated models with numerous sub-models, but to offer a flexible modeling platform with openly accessible codes and data (URL: https://github.com/DrKeHan/DTD) to facilitate model development and implementation in this area.

The remainder of this paper is organized as follows. Section 2 offers a review of relevant literature on DTD and travel choice modeling. We present the proposed DTD models in Section 3. The within-day dynamic network loading model is described in Section 4 and the Appendix. The doubly dynamic models are demonstrated on the Sioux Falls and Anaheim networks in Section 5. Finally, Section 6 offers some managerial insights and concluding remarks.

## 2. Literature Review

In this section we review relevant literature on day-to-day dynamic traffic assignment, bounded rationality, information sharing behavior, and dynamic network loading.

### 2.1. DTD traffic assignment models

DTD models are capable of capturing the transient states disequilibrium and its evolution toward equilibrium induced by construction works, random events and traffic controls, which traditional equilibrium models cannot adequately describe. Watling and Hazelton (2003) state that DTD models are flexible to accommodate a wide range of behavior rules, level of aggregation, and traffic model types. Two types of DTD models have been studied; namely deterministic and stochastic models. The reader is referred to Cantarella and Watling (2016) and Watling and Cantarella (2013, 2015) for a review of deterministic and stochastic DTD models.

Horowitz (1984) first propose a discrete time DTD deterministic models for a two-link network based on the system-optimal principle. Friesz et al. (1994) apply a projective algorithm for the approximation of continuous-time deterministic DTD traffic evolution process. Nagurney and Zhang (1997) also form a projected dynamical system and applied Euler's method to solve it. He et al. (2010) study the continuous-time deterministic approach and the DTD traffic evolution process towards dynamic user equilibrium. On the other hand, stochastic DTD models are suited for modeling traffic variability observed in real-world networks, and are able to represent both transient states and steady-state fluctuations. Cantarella and Watling (2016) present a general stochastic DTD process considering travelers' habits in route choice behavior. Xiao and Lo (2016) consider a microscopic, stochastic process model with departure time choice evolution incorporating information sharing via social media. Wang et al. (2016) propose approximate models for DTD traffic evolution using sensitivity analysis of the (static) network loading process. Rambha and Boyles (2016) propose a stochastic day-to-day dynamic route choice model and then present an average cost Markov decision process to minimize the total network travel time by dynamic pricing. Bifulco et al. (2016) develop a DTD DTA model to capture the effects of advanced traveler information system to departure time choices under recurrent network conditions. Guo and Szeto (2018) propose a DTD dynamic system to model the decision-making processes of travelers and transportation authority as well as their interactions under both private and public transport system. Watling and Hazelton (2018) propose asymptotic approximations to study the transient process in DTD traffic dynamics by relying solely on the knowledge of the equilibrium state without the need for simulation.

The aforementioned DTD literature focuses on the learning and decision making of travelers, with simplified, static within-day traffic models. Only a few studies consider both within-day and day-to-day traffic dynamics; i.e. doubly-dynamic traffic assignment (Cantarella and Astarita 1999; Balijepalli et al., 2007; Friesz et al, 2011; Szeto and Jiang, 2011). These doubly-dynamic DTA models are applied to small traffic networks and only the route choice behavior is considered. The doubly-dynamic DTA model with simultaneous route and departure time choices aims to bridge this gap in the literature.

## 2.2. Bounded rationality in traffic assignment

Conventional user equilibrium type models are based on the behavioral assumption that travelers make rational choices by aiming to minimize their experienced costs. In reality, however, travelers do not always follow the least costly alternative, a phenomenon coined "bounded rationality" (BR) (Simon, 1957; Mahmassani and Chang, 1987). This is corroborated by empirical studies and experiments (Avineri and Prashker, 2004; Zhu and Levinson, 2012). A growing literature on BR in traffic modeling has linked this behavioral mechanism to traffic equilibrium analysis. Since the work of Mahmassani and Chang (1987), BR-based user equilibrium models have been studied as simulation-based DTA (Mahmassani and Jayakrishnan, 1991; Hu and Mahmassani, 1997; Mahmassani and Liu, 1999; Mahmassani et al., 2005) and static traffic assignment (Han and Timmermans, 2006; Gifford and Checherita, 2007; Lou et al., 2010; Guo and Liu, 2011; Watling et al., 2018).

BR has also received increased attention in analytical dynamic traffic assignment. Ridwan (2004)

applies fuzzy system theory to study BR in dynamic traffic modeling. Szeto and Lo (2006) propose a dynamic user equilibrium model with boundedly rational route choice behavior, and apply a heuristic route-swapping algorithm to solve the BR-DUE problem. Ge and Zhou (2012) develop a boundedly rational route choice DUE model with indifference band determined endogenously; but no solution method is provided. Han et al. (2015) propose a BR-DUE model that simultaneously captures route and departure time choices. Solution existence and characterization, as well as three different computational methods are proposed by the authors.

A few studies have adapted the BR notion to DTD models. Guo and Liu (2011) is among the first to propose boundedly rational DTD models with a static, route choice within-day component. The convergence towards a BR user equilibrium and its stability property are later comprehensively analyzed by Di et al. (2015) and Ye and Yang (2017). Guo et al., (2017) investigate a different type of BR-based DTD models with departure time choice at a single bottleneck. Most of existing BR-based DTD models have a static within-day component, and the doubly dynamic ones only focus on departure time choice at a single bottleneck. There does not exist any BR extension of the doubly dynamic DTD models with SRDT choices. Furthermore, all these BR-based DTD dynamics are restricted to deterministic choice models, relying primarily on projective or route-swapping type dynamics (Nagurney and Zhang, 1997; Smith, 1984) assuming complete and perfect travel information.

In this paper, we propose the first boundedly rational doubly dynamic DTA model with SRDT choices. The BR is incorporated in the random utility component of the choice model, which allows travelers' random perception errors to interact with the indifference band in the formulation of the DTD model.

## 2.3. Information sharing and collection behavior in DTD models

Travelers make route and departure time choices based on their perceived travel costs, which directly rely on the levels of availability, relevance, and reliability of travel information they receive. Traditional traffic assignment models tend to assume that travelers have complete information of all other alternatives (Cascetta and Cantarella, 1993; Cantarella and Cascetta, 1995; Watling and Hazelton, 2003; Watling and Cantarella, 2013; Cantarella and Watling, 2016), which is an idea situation compared to a real-world traffic system. On the other hand, the emergence of advanced traveler information system (ATIS) (e.g. Li et al., 2018) as well as social media platforms offer travelers opportunity to perceive parts of the traffic network beyond what they experience on a daily basis, thereby affecting their daily travel choices to a considerable degree. Properly representing the mechanisms of information dissemination and collection thus becomes a crucial part of traffic assignment modeling.

While information incompleteness can be partially taken into account using the notion of perception error of travel costs in a stochastic assignment framework, it is still a meaningful and important undertaking to explicitly model different levels of information availability and sharing behavior. Iryo (2016) develops a deterministic DTD model that explicitly incorporates individual information collection behavior in their daily adjustment. This is done at a microscopic (agent-based) level. Under some restrictive assumptions (e.g. a single user group, among others), the agent-based dynamic is aggregated to derive the macroscopic counterpart as an ordinary differential equation. Xiao and Lo (2016) propose a general framework for DTD commuters' departure time choice, which investigated the influence of friends' information sharing by social media. The day-to-day learning process is modeled with a Bayesian learning theory. Wei et al., (2016) propose a microscopic route choice behavior based on information shared from other travelers, before converting it to a macroscopic traffic flow model, which resembles Smith's proportional route-swapping mechanism (Smith, 1984). Zhang et al. (2018) study the effects of travel information collected from friends on commuters' DTD route choice adjustment based on the cumulative prospect theory. All these DTD models approach information sharing from an agent-based (i.e. microscopic) perspective, which offers valuable insights at individual and system levels. However, there is a lack of generalizable and computationally efficient macroscopic modeling counterpart that is immediately suitable for simulating large-scale dynamic traffic networks. In this paper, we aim to address this issue by proposing an intuitive mathematical model to incorporate information sharing behavior while retaining a computationally tractable form for large-scale simulations.

## 2.4. Dynamic network loading for within-day modeling

As mentioned in the introduction, the majority of DTD models focus on developing behaviorally sound travel choice models while simplifying the within-day component by either resorting to static flow representation, or employing relatively simple dynamic traffic flow models in small networks. For the within-day dynamics, this paper employs a *dynamic network loading* (DNL) procedure based on the Lighthill-Whitham-Richards fluid dynamic model (Lighthill and Whitham, 1955; Richards 1956). This macroscopic perspective of dynamic traffic flow is chosen here not only for its widely recognized capabilities of capturing realistic dynamic traffic network phenomena including shock wave, physical queues, and vehicle spillback, but also for its consistency with the macroscopic travel choice and information sharing models inherent in the day-to-day component. The LWR type kinematic wave model has been widely applied to DTA problems (Han el al., 2015; Garavello et al., 2016; Bliemer et al., 2017), sometimes in its discrete or variational forms such as the cell transmission model (Daganzo, 1994;1995), link transmission model (Yperman et al., 2005), and double-queue model (Osorio et al., 2011). We refer the reader to Nie and Zhang (2005), Garavello et al. (2016) and Han et al. (2016) for a review of relevant literature and computational examples. Individual travelers' route and departure time choices will be represented in a macroscopic way as path departure rates, and the within-day traffic dynamics amount to the DNL procedure, which predicts the corresponding travel costs including travel time and arrival penalties (Friesz et al., 1993).

## 3. Day-to-Day Dynamic Network Models

The proposed DTD modeling framework is comprised of DTD learning and travel choice models, and a within-day dynamic network loading (DNL) model. As illustrated in Figure 1, the DTD model consists of two parts: formulation of perceived travel costs and SRDT choice model. The former presents two different models, one of which involves the information sharing behavior; the latter has three versions, one of which is the proposed BR-based choice model. As the travel cost perception and SRDT choice models are sequential, in total we have six different DTD models.

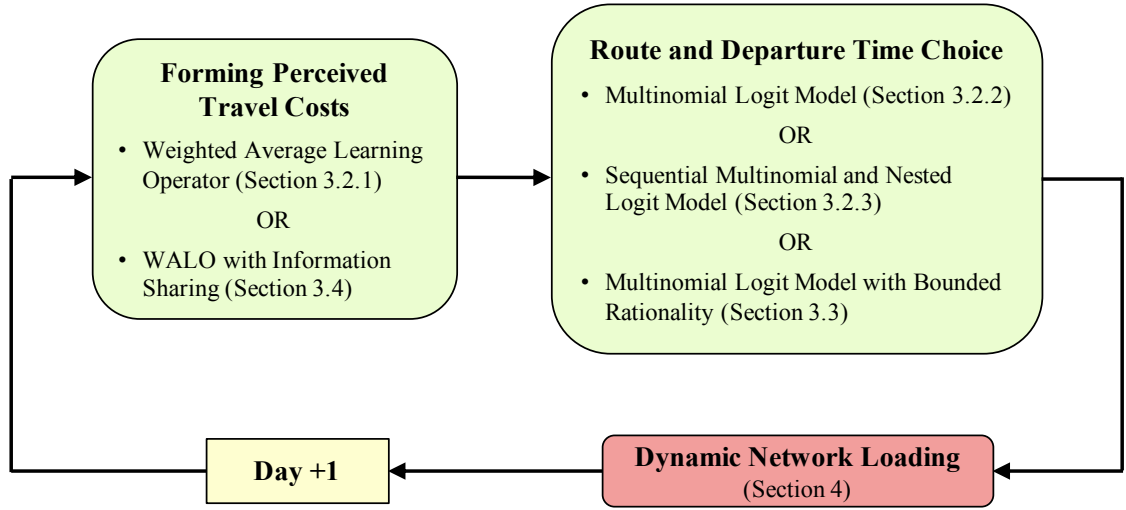

Figure 1. Structure and logic flow of the proposed DTD models.

## 3.1. Notation and essential background

We begin by listing the key notations employed in this paper.

**Parameters/variables**

- $t$: Departure time window
- $s$: Within-day time parameter[1]
- $\tau$: Day-to-day time parameter
- $r$: Route taken by travelers
- $w$: Origin-destination (O-D) pair
- $d^w$: Demand between O-D pair $w$
- $f_{(r,t)}(\tau)$: Departure volume along $r \in R$ at departure window $t$ on day $\tau$
- $\bar{C}_{(r,t)}(\tau)$: Perceived cost for route $r \in R$ and departure time $t$ on day $\tau$
- $C_{(r,t)}(\tau)$: Experienced (actual) cost for route $r \in R$ and departure time $t$ on day $\tau$

**Sets**

- $T$: Set of within-day departure windows
- $R$: Set of all routes in the network
- $W$: Set of all O-D pairs in the network
- $R^w$: Set of routes between O-D pair $w$

**Remark.** This paper employs a route-based assignment approach, as opposed to link-based models (e.g., Ran and Boyce, 1996; He et al., 2010; Long et al., 2018), which means that the route set for each O-D pair is pre-defined. We use a route set generation scheme following Friesz et al. (1992, 1993). Specifically, we apply the Frank-Wolfe algorithm to generate a priori route sets, by making the network increasingly congested through uniformly scaling up the O-D demand matrix, and saving route information produced by the F-W algorithm. This method is intrinsically similar to the well-known technique of column generation as they only calculate routes as they are used. Route sets generated in this way encapsulate essential network information such as O-D demand table, network topology and interactions of O-D flows through congestion; they are arguably more reasonable than others such as those produced by k-shortest-path algorithms, which only consider free-flow link travel times with no consideration of the distribution of O-D demands or their interactions on the network.

---

[1] Here, the parameter $t$ represents departure time window (e.g. 8:00-8:30), while $s$ is on a smaller time scale (e.g. 15 seconds) which is used to represent traffic dynamics in the dynamic network loading procedure discussed in Section 4.

**3.2. SRDT DTD DTA model with imperfect information**

In this paper we invoke the notion of perceived travel cost to account for travel experiences that have been accumulated over the course of the daily congestion game. Throughout the rest of the paper, we stipulate that travelers make choices on route $r$ and departure time $t$.

*3.2.1. Formulation of perceived travel cost*

To this end, we apply the weighted average learning operator (Cascetta 1989; Ouyang, 2007) for calculating the perceived travel cost, denoted $\bar{C}_{(r,t)}(\tau)$ for route $r$ and departure window $t$:

$$\bar{C}_{(r,t)}(\tau) = \frac{1}{s(\lambda)}\left(C_{(r,t)}(\tau-1) + \lambda C_{(r,t)}(\tau-2) + \lambda^2 C_{(r,t)}(\tau-3) + \cdots + \lambda^{M-1}C_{(r,t)}(\tau-M)\right) \quad (1)$$

$$\forall r \in R^w, t \in T$$

where the parameter $\lambda \in (0,1)$ and its powers represent the weights of past days' experienced costs; earlier trips are envisaged to have less influence on the present travel choices, and therefore carry less weight in Equation (1). $M$ is the number past days that influences present day's decision. $1/s(\lambda)$ is a normalization factor where $s(\lambda) = \sum_{i=1}^{M}\lambda^{i-1}$. Equation (1) specifies the dynamics for the perceived cost associated with the choice pair $(r, t)$.

In the following two sections, we present two different choice models based on such perceived costs.

*3.2.2. Multinomial Logit choice model (Base Model I)*

In the first choice model, we treat each alternative as a route-and-departure-time pair $(r,t) \in R^w \times T$, and the perceived costs defined in (1) as the disutility of this alternative. Following the random utility theory, we define the expected travel cost as the sum of the perceived cost and a random observation error term:

$$\hat{C}_{(r,t)}(\tau) = \bar{C}_{(r,t)}(\tau) + \epsilon_{(r,t)}^w \quad \forall r \in R^w, t \in T \quad (2)$$

In case the opposite of the error terms, $-\epsilon_{(r,t)}^w$, are independent and identically Gumbel distributed, the probability of choosing $(r,t)$ is given by the multinomial Logit model:

$$P_{(r,t)}^w(\tau) = \Pr\left\{\bar{C}_{(r,t)}(\tau) + \epsilon_{(r,t)}^w \leq \bar{C}_{(r',t')}(\tau) + \epsilon_{(r',t')}^w \quad \forall (r',t') \neq (r,t)\right\}$$
$$= \frac{\exp\left(-\theta \bar{C}_{(r,t)}(\tau)\right)}{\sum_{(r',t')} \exp\left(-\theta \bar{C}_{(r',t')}(\tau)\right)} \quad \forall r \in R^w, t \in T \tag{3}$$

where $\theta > 0$ is the scale parameter of the Gumbel distribution. Given such probabilities, the departure volumes can be calculated as

$$f_{(r,t)}(\tau) = d^w \cdot P_{(r,t)}^w(\tau) \quad \forall r \in R^w, t \in T \tag{4}$$

We call the model (1), (3) and (4) Base Model I.

By assuming that the departure time choice and route choice are independent, Base Model I, conforms to the conventional notion of SRDT choices in dynamic traffic assignment (Friesz et al., 1993), where travelers' collective choices are represented by the macroscopic quantity of route departure rates $h_r(t)$, where $r$ and $t$ denote route and departure time, respectively. The fact that $h_r(t)$ can be any square-integrable function (or a bounded vector in a discrete-time setting) as long as the demand conservation constraint is satisfied suggests that the departure time and route choices are independent. For this reason, we treat Model Base I as a natural extension of the SRDT notion widely studied in the literature (e.g. Friesz et al. 1993; Szeto and Lo, 2004; Han et al., 2015), when a stochastic choice model based on the random utility theory is employed.

### 3.2.3. Sequential choice model with multinomial and nested Logit models (Base Model II)

Despite the strong ties to other notions of SRDT choices in the literature, Base Model I has some limitations. For example, in reality travelers may consider departure time and route choices sequentially, rather than independently. Furthermore, the multinomial Logit model suffers from the IIA (Independence from Irrelevant Alternatives) assumption, which leads to unrealistically decorrelated perception errors when alternatives have substantial overlap (e.g. routes that share a lot of links).

To address these limitations, in the second choice model, we assume that travelers make decisions about departure time and route sequentially. That is, the probability of choosing the pair $(r,t)$ is given by

$$P_{(r,t)}^w(\tau) = \Pr\{\text{choose departure window } t\} \cdot \Pr\{\text{choose route } r \mid \text{departure window } t\}$$
$$= P_t^w(\tau) \cdot P_{(r|t)}^w(\tau) \quad \forall r \in R^w, t \in T \tag{5}$$

In particular, a traveler makes a departure-time choice based on the following perceived cost associated with the departure time $t \in T$:

$$\bar{C}_t^w(\tau) \doteq \frac{1}{|R^w|} \sum_{r \in R^w} \bar{C}_{(r,t)}(\tau) \quad \forall t \in T \tag{6}$$

Equation (6) means that the perceived cost of departure time window $t$ is the arithmetic mean over all the perceived route costs with the same departure window. The departure-time choice is described by the following multinomial Logit model:

$$P_t^w(\tau) = \frac{\exp(-\theta_1 \bar{C}_t^w(\tau))}{\sum_{t'} \exp(-\theta_1 \bar{C}_{t'}^w(\tau))} \qquad \forall\, t \in T, w \in W \tag{7}$$

where $\theta_1 > 0$ is the scale parameter.

**Remark.** As a variation of Eqn (6), we consider the case where routes with lower disutilities are given higher weights within a departure window. In other words, travelers give more considerations to the better routes when making decisions on their departure times. Specifically, we may consider the following form:

$$\bar{C}_t^w(\tau) \doteq \sum_{r \in R^w} \omega_{(r,t)} \bar{C}_{(r,t)}(\tau), \qquad \omega_{(r,t)} = \frac{1/\bar{C}_{(r,t)}(\tau)}{\sum_{r' \in R^w} 1/\bar{C}_{(r',t)}(\tau)} \qquad \forall\, t \in T \tag{8}$$

which reduces $\bar{C}_t^w(\tau)$ to the harmonic mean of $\{\bar{C}_{r,t}(\tau), r \in R^w\}$:

$$\bar{C}_t^w(\tau) \doteq \frac{|R^w|}{\sum_{r \in R^w} 1/\bar{C}_{(r,t)}(\tau)}, \qquad \forall\, t \in T \tag{9}$$

This variant will be tested later in Section 5.3 in comparison with (6).

Once the departure window $t$ is chosen, the probability of choosing route $r \in R^w$ is expressed by the nested Logit model instead of the multinomial Logit model to correct the IIA assumption and account for the correlations in the perception errors for different route alternatives (Bierlaire and Frejinger, 2005). For this purpose, we employ the Path Size formulation (Ben-Akiva and Bierlaire, 2003) as follows. We define the Path Size as an attribute of each route $r \in R^w$

$$\text{PS}_r = \sum_{a \in r} \frac{L_a}{L_r} \cdot \frac{1}{\sum_{r' \in R^w} \delta_{ar'}} \tag{10}$$

where the route $r$ is expressed as a set of arcs $a$ that it traverses; the lengths of the arc and the route are denoted $L_a$ and $L_r$, respectively. Moreover, $\delta_{ar} = 1$ if $a \in r$ and 0 otherwise. The static attribute $\text{PS}_r$ reflects the level of overlap (number of shared links) between two routes, and the corrected route choice probability reads

$$P_{(r|t)}^w(\tau) = \frac{\exp\left(-\theta(\bar{C}_{(r,t)}(\tau) + \eta \ln \text{PS}_r)\right)}{\sum_{r'} \exp\left(-\theta(\bar{C}_{(r',t')}(\tau) + \eta \ln \text{PS}_{r'})\right)} \qquad r \in R^w \tag{11}$$

where $\eta > 0$ is the weight for the Path Size attribute. Due to space limitation we omit the detailed derivation of the PS-based nested Logit model, and refer the reader to Ben-Akiva and Bierlaire (2003) and Bierlaire and Frejinger (2005). Finally, the departure volumes are given by

$$f_{(r,t)}(\tau) = d^w \cdot P_t^w(\tau) \cdot P_{(r|t)}^w(\tau) \qquad \forall\, r \in R^w, w \in W, t \in T \tag{12}$$

We call the model (1) and (6)-(12) Base Model II.

### 3.2.4. Steady states of Base Models I and II

In this section, we analyze the steady states of the proposed two DTD models, and show that they correspond to relevant notions of stochastic dynamic user equilibrium (SDUE) with SRDT choices.

We begin by defining such equilibria for the choice behavior elaborated in Section 3.2.2 (SRDT-SDUE-I) and Section 3.2.3 (SRDT-SDUE-II).

**Definition (SRDT-SDUE-I).** *A path departure rate vector $\left(f^*_{(r,t)}: r \in R^w, t \in T\right)$ corresponds to a SRDT-SDUE-I solution if it satisfies the following:*

$$f^*_{(r,t)} = d^w \cdot \frac{\exp\left(-\theta C^*_{(r,t)}\right)}{\sum_{(r',t')} \exp\left(-\theta C^*_{(r',t')}\right)}, \quad \forall r \in R^w, w \in W, t \in T \quad (13)$$

where $C^*_{(r,t)}$ is the travel cost associated with route $r$ and departure time $t$, which is given by the DNL sub-model.

**Definition (SRDT-SDUE-II).** *A path departure rate vector $\left(f^*_{(r,t)}: r \in R^w, t \in T\right)$ corresponds to a SRDT-SDUE-II solution if it satisfies the following:*

$$f^*_{(r,t)} = d^w \cdot \frac{\exp(-\theta_1 \bar{C}^{w,*}_t)}{\sum_{t'} \exp(-\theta_1 \bar{C}^{w,*}_{t'})} \cdot \frac{\exp\left(-\theta\left(C^*_{(r,t)} + \eta \ln \text{PS}_r\right)\right)}{\sum_{r'} \exp\left(-\theta\left(C^*_{(r',t')} + \eta \ln \text{PS}_{r'}\right)\right)} \quad \forall r \in R^w, w \in W, t \in T \quad (14)$$

where $\bar{C}^{w,*}_t = \frac{1}{|R^w|} \sum_{r \in R^w} C^*_{(r,t)}$, $C^*_{(r,t)}$ is the travel cost associated with route $r$ and departure time $t$, $\text{PS}_r$ is given in Eqn. (10).

**Proposition (steady states of Base Models I).** *If the DTD process detailed in Base Models I converges, then the steady state is a SRDT-SDUE-I solution. In other words, if*

$$f_{(r,t)}(\tau) = f_{(r,t)}(\tau - 1) = f_{(r,t)}(\tau - 2) = \cdots = f_{(r,t)}(\tau - M), \quad \forall r \in R^w, w \in W, t \in T \quad (15)$$

*then there must hold*

$$f_{(r,t)}(\tau) = d^w \cdot \frac{\exp\left(-\theta C_{(r,t)}(\tau)\right)}{\sum_{(r',t')} \exp\left(-\theta C_{(r',t')}(\tau)\right)}, \quad \forall r \in R^w, w \in W, t \in T \quad (16)$$

Proof. Given (15), we must have that $C_{(r,t)}(\tau) = C_{(r,t)}(\tau - 1) = \cdots = C_{(r,t)}(\tau - M), \forall r \in R^w, t \in T$. Therefore, Eqn. (1) yields $\bar{C}_{(r,t)}(\tau) = C_{(r,t)}(\tau - 1) = \cdots = C_{(r,t)}(\tau - M)$. By definition (4), we have:

$$f_{(r,t)}(\tau) = d^w \cdot \frac{\exp\left(-\theta \bar{C}_{(r,t)}(\tau)\right)}{\sum_{(r',t')} \exp\left(-\theta \bar{C}_{(r',t')}(\tau)\right)} = d^w \cdot \frac{\exp\left(-\theta C_{(r,t)}(\tau)\right)}{\sum_{(r',t')} \exp\left(-\theta C_{(r',t')}(\tau)\right)}$$

Following the same proof, we readily deduce:

**Proposition (steady states of Base Models II).** *If the DTD process detailed in Base Models II converges, then the steady state is a SRDT-SDUE-II solution.*

**Remark.** An important theoretical issue is the convergence of Base Models I & II to their respective equilibrium. However, this is a very difficult problem due to the lack of analytical and provable regularity conditions of the network delay operator. In particular, the uniqueness of the aforementioned SDUEs and the convergence of the proposed DTD models are dependent on the generalized monotonicity of the delay operator (Szeto and Lo, 2004; Friesz et al., 2011; Han et al., 2019), which, in the case of large-scale and arbitrary traffic networks, are difficult to prove. In fact, they are unlikely to hold. Furthermore, the convergence is also dependent on model parameters such as $\theta$ in the Logit model, as well as the initial traffic state. While these theoretical problems cannot be solved in this paper, our numerical results do present insightful findings on the behaviors of the DTD processes. As we clarify at the beginning of this paper, this work is intended for modeling scenarios where a disequilibrium model is required (such as adaptive network traffic control or management, transient congestion), where a flexible tool for large-scale network modeling and simulation is missing.

### 3.3. DTD DTA model with bounded rationality

Building on the Base Model I presented in Section 3.2.2, we further consider the scenario where some travelers may prefer to maintain their previous choices if the expected benefits of switching to alternatives is insignificant. This choice behavior has been well documented as bounded rationality (BR) in many existing studies (see Section 2.2), but none has investigated BR in a doubly dynamic context while incorporating imperfect and incomplete information.

**Remark.** *We choose to combine BR with Base Model I instead of Base Model II for the following reasons. (1) It is mathematically complicated to incorporate the notion of BR in Base Model II, given its sequential choice structure. We will leave this to a future study. (2) The combination of Base Model I + BR offers better generality and transferability to other types of DTD models, such as DTD models with static traffic flow.*

For modeling purposes, the BR is often represented as an indifference band (Mahmassani and Chang, 1987; Han et al., 2015), denoted $\delta \geq 0$, which measures the acceptable difference between the expected cost of a reference choice, and the minimum expected cost among all alternatives. In other words, a traveler will not switch to alternatives on the next day if such a difference is smaller than $\delta$. To quote the taxonomy of Xu et al. (2017), the type of BR choice considered here is status quo-dependent in the sense that travelers' choices are in reference to their choice on the previous day (the status quo) instead of the best (most cost-effective) alternative.

We fix a given choice pair $(r,t) \in R^w \times T$. At an atomic (microscopic) level, the BR choice principle can be described as follows: a traveler with a choice of $(r,t)$ on day $\tau - 1$ does not alter his/her route and departure time choices on day $\tau$ provided that the expected costs satisfy

$$\hat{C}_{(r,t)}(\tau) - \hat{C}_{(r',t')}(\tau) \leq \delta \quad \forall (r',t') \neq (r,t) \tag{17}$$

(17) means that no alternative can offer a gain larger than $\delta$. Therefore, on a macroscopic level, the probability of not changing the travel choice on day $\tau$ (that is, keep choosing $(r,t)$), denoted $\phi_{(r,t)}^w(\tau)$, is:

$$\phi^w_{(r,t)}(\tau) = \Pr\left\{\hat{C}_{(r,t)}(\tau) - \hat{C}_{(r',t')}(\tau) \leq \delta \quad \forall (r',t') \neq (r,t)\right\}$$

$$= \Pr\left\{\bar{C}_{(r,t)}(\tau) - \delta + \epsilon^w_{(r,t)} \leq \bar{C}_{(r',t')}(\tau) + \epsilon^w_{(r,t)} \quad \forall (r',t') \neq (r,t)\right\} \quad (18)$$

$$= \frac{\exp\left(-\theta(\bar{C}_{(r,t)}(\tau) - \delta)\right)}{\exp\left(-\theta(\bar{C}_{(r,t)}(\tau) - \delta)\right) + \sum_{(r',t')\neq(r,t)} \exp(-\theta \bar{C}_{(r',t')}(\tau))}$$

Here, the last equality is established by viewing $\bar{C}_{(r,t)}(\tau) - \delta$ as the disutility associated with the choice $(r,t)$. (18) is seen as an extension of the binary logit model with indifference (Krishnan, 1977).

On the other hand, a traveler with a choice of $(r,t)$ on day $\tau - 1$ choses $(r_1, t_1) \neq (r,t)$ on day $\tau$ if and only if

$$\hat{C}_{(r_1,t_1)}(\tau) < \hat{C}_{(r,t)}(\tau) - \delta \quad \text{and} \quad \hat{C}_{(r_1,t_1)}(\tau) < \hat{C}_{(r',t')}(\tau) \quad \forall (r',t') \neq (r,t) \text{ or } (r_1, t_1) \quad (19)$$

Based on (19) and our calculation (18), it is not difficult to conclude that the choice modeling, among travelers who chose $(r,t)$ on day $\tau - 1$, amounts to a multinomial Logit model where the disutility of $(r,t)$ is revised to be $\bar{C}_{(r,t)}(\tau) - \delta$, and the systematic disutilities of other alternatives remain unchanged. Note that each choice pair $(r,t)$ is associated with such a multinomial Logit model. Therefore, the departure volume for the choice $(r,t)$ on day $\tau$ can be calculated as

$$\begin{aligned} f_{(r,t)}(\tau+1) &= \phi^w_{(r,t)}(\tau) \cdot f_{(r,t)}(\tau) \\ &+ \sum_{(r',t')\neq(r,t)} \frac{\exp\left(-\theta \bar{C}_{(r,t)}(\tau)\right)}{\exp\left(-\theta(\bar{C}_{(r',t')}(\tau) - \delta)\right) + \sum_{(r_1,t_1)\neq(r',t')} \exp\left(-\theta \bar{C}_{(r_1,t_1)}(\tau)\right)} \cdot f_{(r',t')}(\tau) \end{aligned} \quad (20)$$

where the first term on the right-hand side represents the amount of travelers who stick with their previous choice $(r,t)$ due to BR, and the second summation term represents travelers who switch to $(r,t)$ from their previous choices. We use Figure 2 to illustrate (20), where three alternatives, A, B and C are considered. Within each alternative there are two types of flows: sticking with previous choice, which is denoted by the curved arrows and represented as the first term of (20), and switching to other alternatives, which is denoted by the straight arrows and represented as the second term of (20).

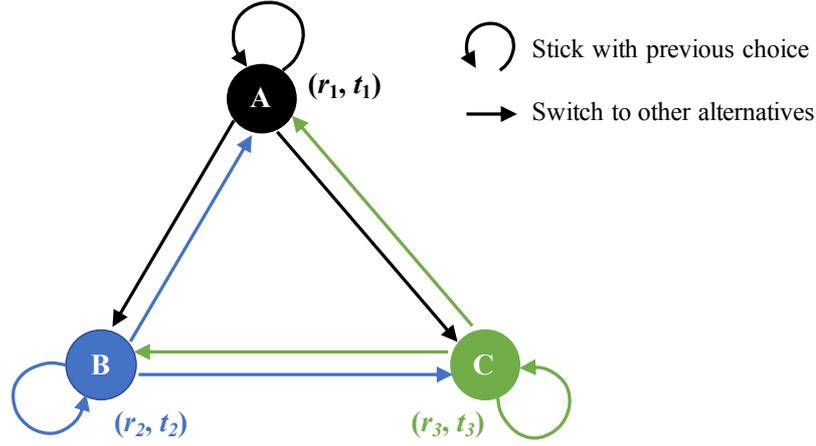

Figure 2. Illustration of the DTD model with bounded rationality.

**Remark.** *In the derivation of the BR DTD DTA model, the key idea lies in the incorporation of the indifference band δ into the random utility model, namely* (17)*,* (18) *and* (19)*. The same modeling framework can be applied to other types of DTD models, such as route choice DTD models with static traffic flow model.*

### 3.4. DTD DTA model with information sharing behavior

In this section, we factor information sharing into travelers' route and departure time decisions. This is achieved by invoking the notion of information reliability, which depends on the number of travelers using each alternative.

We fix a choice pair $(r,t) \in R^w \times T$. The information sharing behavior can be described as follows. Travelers choosing $(r,t)$ on day $\tau-1$ share their experienced travel cost $C_{(r,t)}(\tau-1)$ within a group, which is defined here as the set of travelers between the same O-D pair $w$.[2] Such information will be utilized, with a weight $g^w_{(r,t)}(\tau-1)$, by the rest of the group to inform their own perceived costs. Here we assume that the weight

$$g^w_{(r,t)}(\tau-1) = G\left(\frac{f_{(r,t)}(\tau-1)}{d^w}\right), \qquad \forall (r,t) \in R^w \times T \tag{21}$$

where $G(\cdot)$ is an increasing function that satisfies $G(0) = 0, G(1) = 1$; it is a function of the traveler proportion who use $(r,t)$ on day $\tau-1$. Given such weights, the perceived travel cost on day $\tau$ is given by

$$\bar{C}_{(r,t)}(\tau) = \frac{1}{s(\lambda,g)} \sum_{n=\tau-M}^{\tau-1} g^w_{(r,t)}(n) \cdot \lambda^{\tau-n-1} \cdot C_{(r,t)}(n) \quad \forall (r,t) \in R^w \times T \tag{22}$$

which is adapted from (1) by adding the multiplicative weights. Here, the normalization factor

---

[2] In fact, the group can be arbitrarily defined without affecting the model formulation. Here we treat travellers between each O-D pair as a group for simplicity.

$$s(\lambda, g) = \sum_{n=\tau-M}^{\tau-1} g_{(r,t)}^{w}(n) \cdot \lambda^{\tau-n-1} \tag{23}$$

Intuitively, model (22) indicates that the cost of certain pair $(r,t)$ perceived by the group is accumulated from the past experienced costs; and the more travelers use $(r,t)$ on certain day, the more significantly their collective experience affects the overall perception of $(r,t)$. This is a reasonable assumption as the reliability of travel information depends on the volume of travelers who report it. Moreover, the functional form of $G$ is likely to be non-linear and may be piece-wise defined.

One choice for the weighting function is $G(x) = x^n$ where $x \in [0,1]$, $n > 0$ (other choices of such a function will be tested later in Section 5.3). Here, the argument $x$ represents the percentage of travelers within a group (O-D pair) who chose $(r,t)$ on a given past day. Our stipulation that $G(x)$ is monotonically increasing reflects the reasonable assumption that more travelers choosing $(r,t)$ leads to higher reliability of their reported information, which receives larger weight in forming individuals' perceptions towards $(r,t)$. When $n = 1$, the weights are proportional to the corresponding percentages. For $n > 1$, the travelers are prone to information reported by larger crowds, and tend to ignore experienced costs reported by smaller crowds. And such a tendency will be intensified as $n$ becomes larger. For $0 < n < 1$, the tendency is reversed in the sense that even a small crowd could influence the perception to a degree not significantly lower than what a much larger crowd can achieve. See Figure 3. In the case where $n = 0$, the model reduces to the case with complete travel information (1). Therefore, the parameter $n$ can be treated as a simplified representation of the strength of communication among travelers. Different values of $n$ correspond to different behavioral situations, and are worthy of further investigation beyond this paper. More numerical insights regarding $n$ will be provided in Section 5; in particular, $n$ has an impact on the daily oscillation of network traffic and travelers' perceptions, and such impact is case dependent.

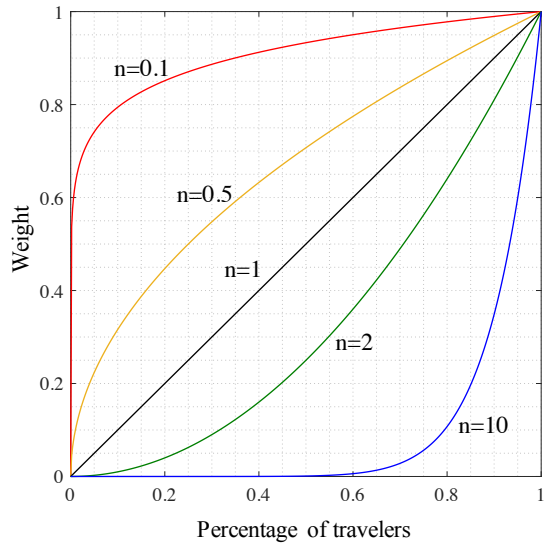

Figure 3. Function $G(x) = x^n$ that expresses the weight (reliability) of experienced information as a function of percentage of travelers that chose $(r,t)$.

The proposed macroscopic information sharing behavior is articulated at the travel cost perception level. This may be immediately applied in conjunction with the travel choice models discussed in Sections 3.2 and 3.3; see Figure 1.

## 4. Within-Day Dynamic Network Loading Model

Section 3 presents a learning and decision-making framework for daily adjustment of travel choices in terms of route and departure time. In other words, it illustrates how the experienced travel costs on day $\tau - 1$, together with the perceived costs accumulated from past experience, can jointly affect travelers' decisions on the next day $\tau$, that is, the day-to-day dynamics. The within-day dynamics, on the other hand, determines the physical states of the traffic network and experienced travel costs on day $\tau$, which allows the process to continue towards the next day (see Figure 1).

The within-day component of the proposed doubly-dynamic model is, in effect, a dynamic network loading (DNL) model (Friesz et al., 1993). The DNL procedure aims at describing and predicting the dynamic evolution of traffic flows and congestion on a road network consistent with traffic flow theory and established route and departure time choices of travelers. In view of the models proposed in Section 3, the main purpose of DNL is to numerically evaluate the experienced cost $C_{(r,t)}$ for all $r \in R^w$, $w \in W$, $t \in T$ with given departure profile $f_{(r,t)}$, $r \in R^w$, $w \in W$, $t \in T$.

### 4.1. Relevant notations for the DNL model

We let $s = 1, 2, 3, ...$ be the discrete time steps with step size $ds$. A network is represented as a directed graph consisting of links and nodes. The following additional notations are introduced to facilitate our presentation. Since we only discuss within-day dynamics here, the day label $\tau$ is dropped throughout the section.

$f_r(s)$: Route departure rate along $r \in R^w$ at time $s$

$f(s)$: Set of route departure rates $f(s) = (f_r(s): r \in R)$ at time $s$

$\text{TT}_r(s)$: Travel time along route $r$ with departure time $s$

$C_r(s)$: Travel cost along route $r$ with departure time $s$

$dt$: Duration of a single departure window

$ds$: Time step size for the dynamic network loading

The input of the DNL problem is the set of route departure rates $f(s) = (f_r(s): r \in R)$. Through computations involving link dynamics, junction dynamics, link delay and path delay, the DNL calculates path travel times (path delays) as $\text{TT}_r(s)$ for route $r \in R$ and departure time $s$. The modeling of departure time choice requires the specification of generalized travel cost with arrival penalties:

$$C_r(s) = \alpha \cdot \text{TT}_r(s) + \beta \cdot \text{EP}_r(s) + \gamma \cdot \text{LP}_r(s) \quad (24)$$

where the early arrival penalty $\text{EP}_r(s)$ and late arrival penalty $\text{LP}_r(s)$ are part of the travel cost, and are relative to a target arrival time TA. $\alpha, \beta, \gamma$ are positive parameter to balance the weights of travel time, and early and late penalties. We reasonably set $\beta < \alpha < \gamma$ to reflect the different values of time (Small, 1982).

We note the important difference between route departure rate $f_r(s)$ and departure volume $f_{(r,t)}$. The former is associated with the DNL model and the latter is invoked in the DTD choice model in Section 3. Moreover, $f_r(s)$ represents flow (unit: vehicle/unit time) defined for every time step $s$, while $f_{(r,t)}$ represents traffic volume (unit: vehicle) in a given departure window $t \in T$.

In the LWR-based dynamic network loading, the time step size $ds$ should satisfy the Courant-Friedrichs-Lewy numerical stability condition (LeVeque, 1992), which means that it should be no longer than the free-flow time on any link in the network. In this paper, we deliberately make the duration of the departure window $dt$ (e.g. 15 min) significantly larger than the time step size $ds$ (e.g. 15 s) to reduce departure time uncertainties. To reconcile both time scales, we propose the following procedure to convert the departure volumes $f_{(r,t)}$ to the route departure rates $f_r(s)$ and then to the experienced costs $C_{(r,t)}$.

1. Given the departure volumes $f_{(r,t)}$ for all $(r,t) \in R^w \times T$, we compute the average departure rate as $f_{(r,t)}/dt$ and set

$$f_r(s) = \frac{f_{(r,t)}}{dt} \quad \text{for all } s \text{ within the departure window } t \qquad (25)$$

2. Perform the DNL procedure (see the Appendix for details) with $f_r(s), r \in R$ given by (25), and obtain the path travel costs $C_r(s) \; \forall r \in R$.

3. Average the costs $C_r(s)$ over the departure window to obtain the costs $C_{(r,t)}$:

$$C_{(r,t)} = \frac{ds}{dt} \cdot \sum_{s \in t} C_r(s) \quad \forall \, r \in R^w, t \in T \qquad (26)$$

### 4.2. The dynamic network loading procedure

In deriving the DNL model, we employ the variational formulation of the Lighthill-Whitham-Richards model known as the Lax-Hopf formula (Han et al., 2016), which enables us to formulate the DNL model as a system of differential algebraic equations (DAEs) and solve the DNL problem with computational efficiency. As these materials are already documented in Han et al. (2019), we omit technical details here and only present the discretized DAE in the Appendix.

## 5. Numerical Case Studies

In this section, we illustrate the proposed DTD models on two test networks: the Sioux Falls network with 528 O-D pairs, 30,000 trips and 6,180 routes, and the Anaheim network with 1,406 O-D pairs, 30,000 trips and 30,719 routes; see Figure 4. Unlike existing literature, which mainly uses small networks and static traffic flow model to illustrate the main features of the DTD learning and decision making on the demand side, we use these reasonably large-scale networks to demonstrate the complexity arising from the interaction between DTD and within-day traffic dynamics, while demonstrating that the proposed models' capability to be readily applied for problems of realistic sizes. The within-day DNL procedure can capture shock wave, spillback, and the inter-link propagation of location congestion in space and time, which is an important feature of the DTD doubly dynamic model.

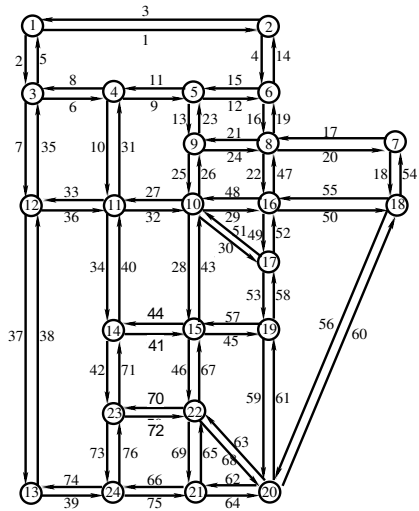 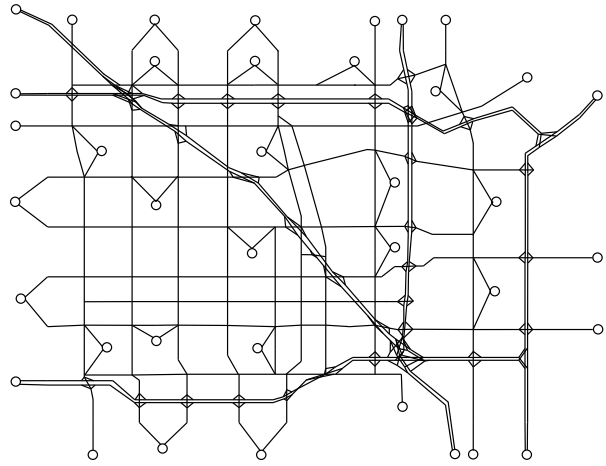

Sioux Falls network
(24 nodes, 76 links, 24 zones)

Anaheim network
(416 nodes, 914 links, 38 zones)

Figure 4. The test networks

The simulation horizon is a five-hour morning commuting period, split into 20 departure time windows (15 min each). In addition to analyzing the long-term behavior of the doubly DTD dynamics towards equilibrium, we also investigate the effect of possible local network disruptions by reducing the capacities of some links before restoring them at the end of the disruption period. A range of DTD models proposed in this paper will be discussed and compared using sensitivity and scenario-based analyses.

Throughout the numerical tests, the travel cost structure follows that of (24) with the following parameters: in-vehicle value of time $\alpha = 1$, early arrival value of time $\beta = 0.8$, late arrival value of time: $\gamma = 1.8$. The unit of the travel cost is seconds.

### 5.1. Long-term behavior of the doubly dynamic models

We begin by examining the long-term behavior of the proposed models by performing a sensitivity analysis on the model parameters. Note that the purpose of the analysis is not to seek any form of dynamic user equilibria, which are often viewed and studied as the asymptotic states of DTD dynamics. Rather, we aim to capture realistic user and traffic behaviors and their interactions, noting that the proposed doubly dynamics may not converge to their steady states due to the highly complex and non-monotone delay operators (Han et al., 2015) associated with the dynamic network loading. Indeed, our study is partially driven by the expectation that idealized equilibria may not exist in real-world traffic systems.

We first test four variants of the proposed DTD models:

1. Base Model I: multinomial Logit model (Section 3.2.2);
2. Base Model II: sequential-decision model with mixed multinomial and nested Logit model (Section 3.2.3);
3. Base Model I with bounded rationality (BR); and
4. Base Model II with information sharing (IS)

The level of daily oscillation is measured by the relative gap:

$$\text{Relative Gap:} \left( \frac{\sum_{r \in R} \sum_{t \in T} \left( f_{(r,t)}(\tau) - f_{(r,t)}(\tau-1) \right)^2}{\sum_{r \in R} \sum_{t \in T} \left( f_{(r,t)}(\tau-1) \right)^2} \right)^{1/2} \quad (27)$$

which represents the relative change of the departure flows in two consecutive days.

### 5.1.1. Long-term behavior on the Sioux Falls network

In Figure 5 and Figure 6, we show the relative gaps for the four model variants with $M = 3$ and $M = 6$, respectively. Other relevant parameters are:

Base Model I: $\lambda = 0.7$, $\theta = 0.004$

Base Model II: $\lambda = 0.7$, $\theta = \theta_1 = 0.004$, $\eta = 400$

Base Model I + BR: $\lambda = 0.7$, $\theta = 0.004$, $\delta = 400$

Base Model II + IS: $\lambda = 0.7$, $\theta = \theta_1 = 0.004$, $\eta = 400$, $G(x) = x^n$

Here $M$ denotes the number of past days that influence present day's perception. Given the learning process (1), which resembles a moving average, it is expected that $M$ has a smoothing effect on the time series (in $\tau$) of the perception of different alternatives, and therefore larger $M$ tends to reduce the daily variations of path flows, hence the relative gaps. This is indeed the case by comparing Figure 5 ($M = 3$) and Figure 6 ($M = 6$). In addition, larger $\lambda$ means that past days' experience carries less weight, which has a similar effect as small $M$. For this reason, in what follows we do not vary the value of $\lambda$, but to only consider different values of $M$.

We further observe from Figure 5 that: (i) Base Model I with bounded rationality has lower relative gaps than without, which is expected because of travelers reluctance to switch choices; (ii) Base Model II with information sharing has larger gaps than without, which highlights the uncertainties in decision making due to lack of complete information (also see Figure 8).

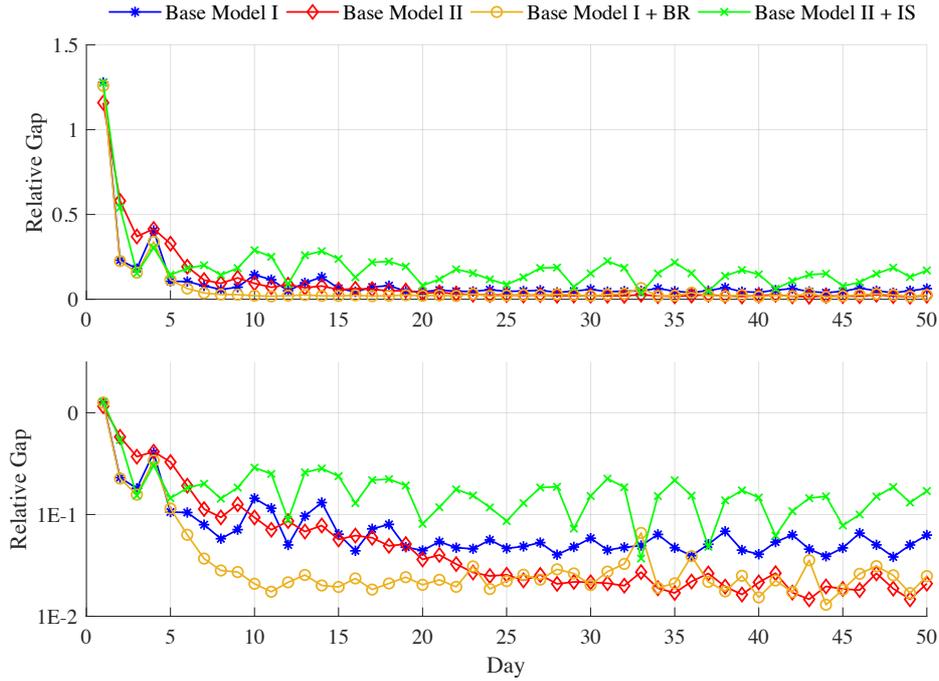

Figure 5. Relative gaps in 50-day simulation ($M = 3$) based on the Base Model I, Base Model II, Base Model I with bounded rationality (BR), and Base Model II with information sharing (IS). The lower figure displays the logarithmic scale.

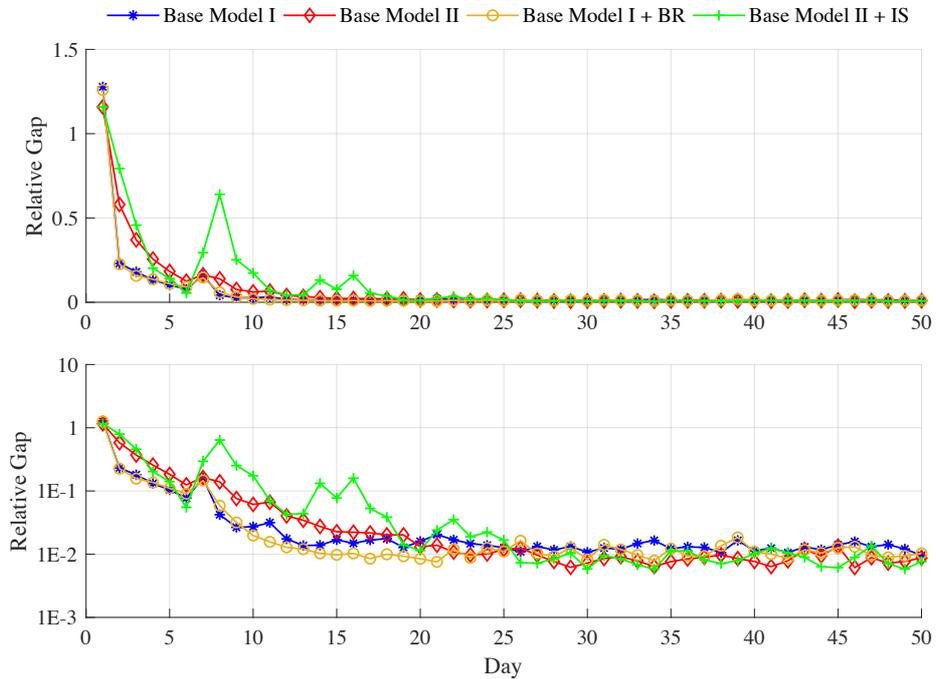

Figure 6. Relative gaps in 50-day simulation ($M = 6$) based on the Base Model I, Base Model II, Base Model I with bounded rationality (BR), and Base Model II with information sharing (IS). The lower figure displays the logarithmic scale.

Figure 7 compares the Base Model I + BR model with different indifference bands $\delta = 0, 200, 400$ and 800. We see a decreasing trend of the relative gaps as $\delta$ increases, which means that when the travelers are willing to accept larger cost differences, their route and departure time choices undergo less daily oscillation.

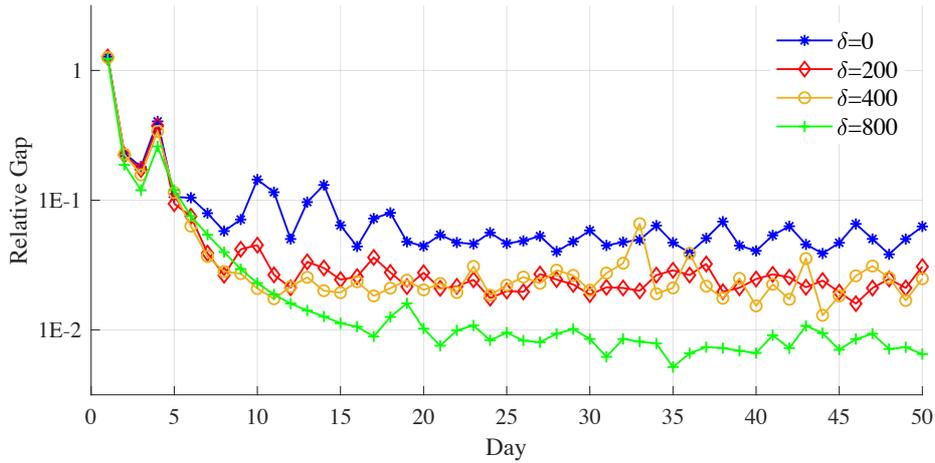

Figure 7. Comparison in terms of relative gaps (in logarithmic scale) of the Base Model I + BR with different values of the indifference band $\delta$.

We use Figure 8 to illustrate the impact of information sharing strength, i.e. $n$ in $G(x) = x^n$, on the variations of travel choices. It can be seen that $n = 2$ yields large relative gaps, which means that ignoring information reported by small crowds, however relevant they may be, tends to create much uncertainties in the decision-making process. However, we note that this is likely the case in a real-world situation, where information provided either by individual travelers via social platforms or by a centralized information sharing entity (e.g. ATIS) tends to focus on those popular choices. Moreover, the other three cases ($n = 0, 0.5, 1$) seem to have reached the same level of relative gaps after 30 days.

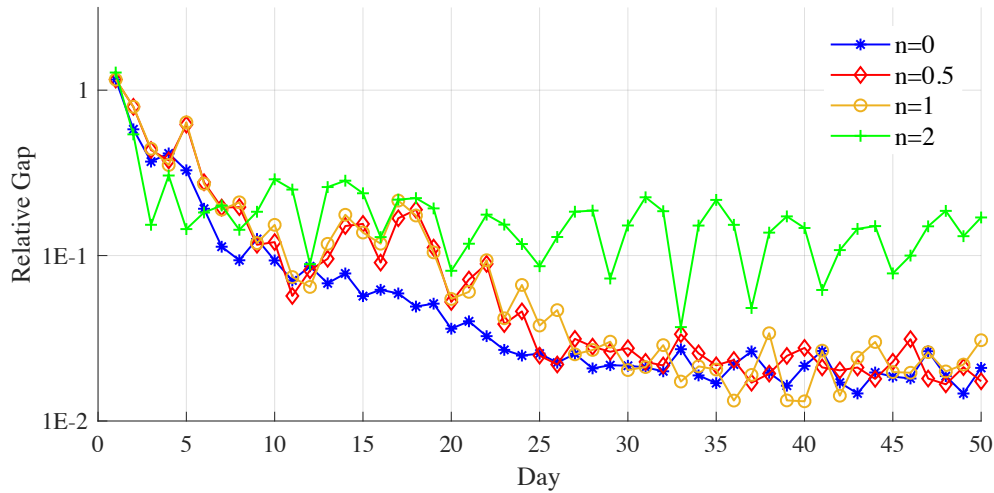

Figure 8. Comparison in terms of relative gaps (in logarithmic scale) of the Base Model II + IS with different values of $n$ in the weighting function $G(x) = x^n$.

### 5.1.2. *Long-term behavior on the Anaheim network*

We perform similar analysis on the Anaheim network. Figure 9 compares the relative gaps produced by Base Model I with different scale parameters $\theta = 0.001, 0.002, 0.003$. We note that larger $\theta$ means that travelers are more sensitive to the cost different between alternatives. We can see that the relative gaps correspond to higher values of $\theta$, which means that higher sensitivity towards the perceived cost difference causes traffic to constantly switch routes and departure times.

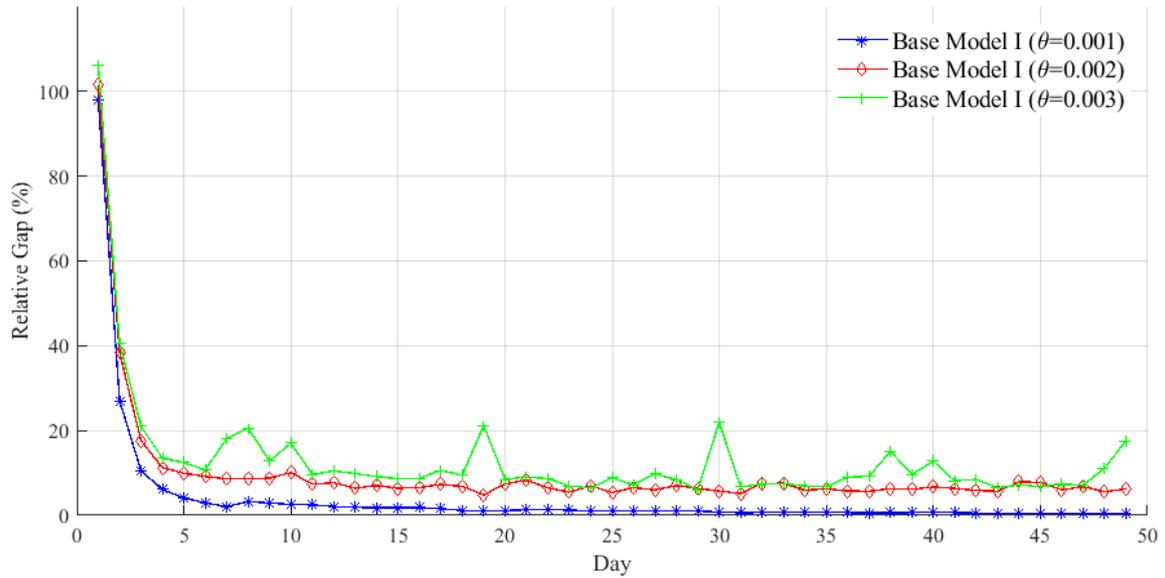

Figure 9. Relative gaps on the Anaheim network produced by Base Model I ($M = 6$).

Next, we examine the effect of information sharing on network performance. We simulate the network dynamics using Base Model II without information sharing (IS), as well as with IS ($n = 1, 2$). The relative gaps and total network costs are shown in Figure 10. It is interesting to see that Base Model II, which is assuming complete traffic information, actually yields the largest relative gap and also highest network total cost compare to the case with incomplete information. In particular, the sums of network costs over the period $[10, 50]$ (day) are:

- $3.4451 \times 10^9$ (Base Model II),
- $3.4400 \times 10^9$ (Base Model II + IS $n = 1$), and
- $3.4393 \times 10^9$ (Base Model II + IS $n = 2$).

This suggests that limiting access to information of certain alternatives could in fact stabilize traffic and reduce congestion associated with daily SRDT choice switches. This can be viewed as a generalized Braess paradox where information transparency is working against the network performance.

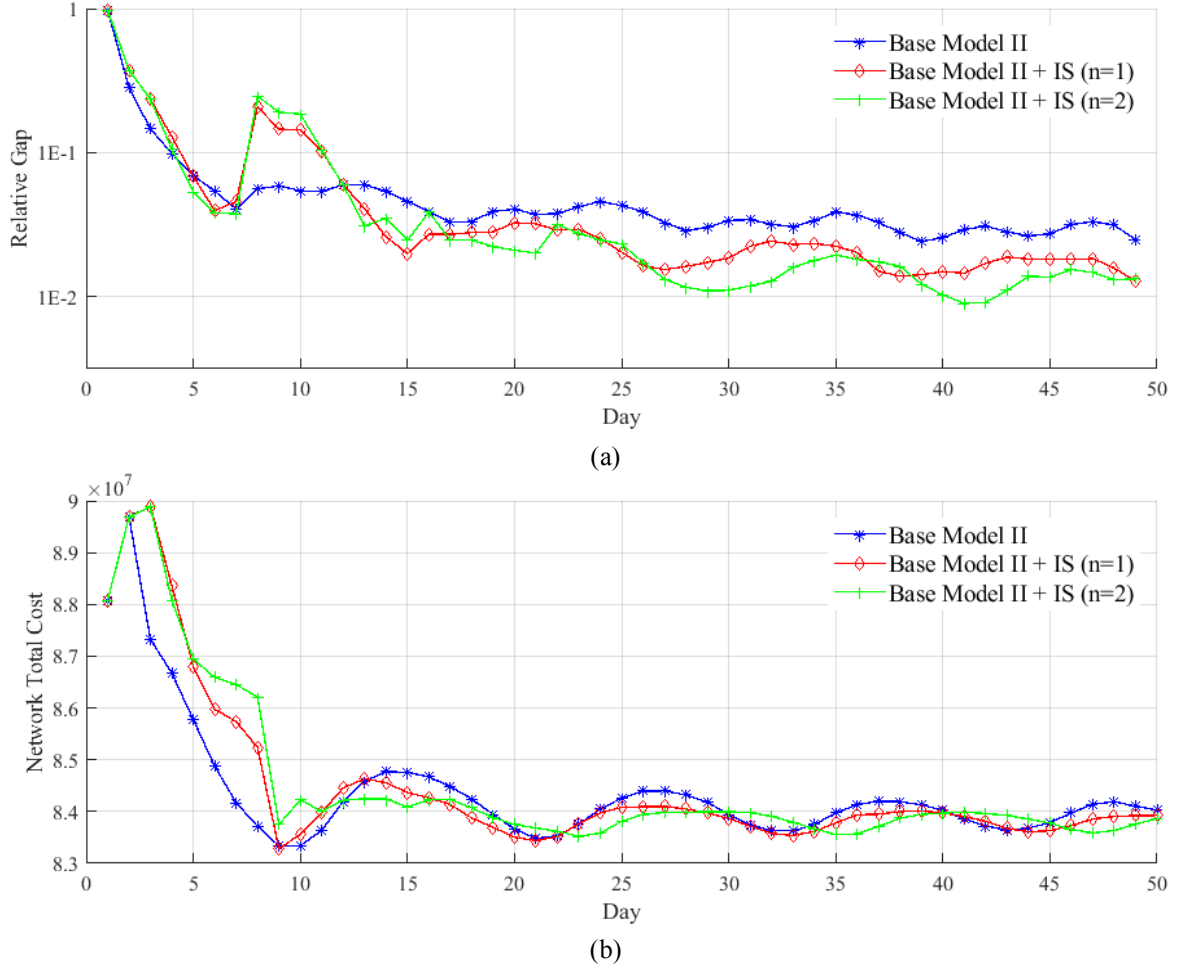

Figure 10. Relative gaps (in logarithmic scale) and network total costs corresponding to Base Model II ($n = 0$) and Base Model II with information sharing ($n = 1, 2$). The other parameters are chosen to be ($\theta = \theta_1 = 0.001, M = 6, \eta = 600$).

We consider a given O-D pair $w \in W$, which has a route set ranging from #39-#219. For each day of simulation, we calculate the average perceived travel cost corresponding to each departure window $t \in \{1, 2, \ldots, 20\}$ as

$$\bar{C}_t^w(\tau) \doteq \frac{1}{|R^w|} \sum_{r \in R^w} \bar{C}_{(r,t)}(\tau)$$

Figure 11 shows the daily evolution of such average perceived costs for four cases: Base Model II + information sharing with $n = 0, 0.5, 1, 2$, respectively. Note that $n = 0$ corresponds to the complete information case. It can be seen that the lowest perceived costs are concentrated between departure windows 8 and 11 in all four cases. However, the difference lies in the departure windows 12-15, where $n = 0$ yields clear daily oscillations of the perceived cost, while such oscillations diminish as $n$ gets larger. This means that the lack of complete information tends to have a smoothing effect on the daily profile of the perceived cost, which conforms with Figure 10(a).

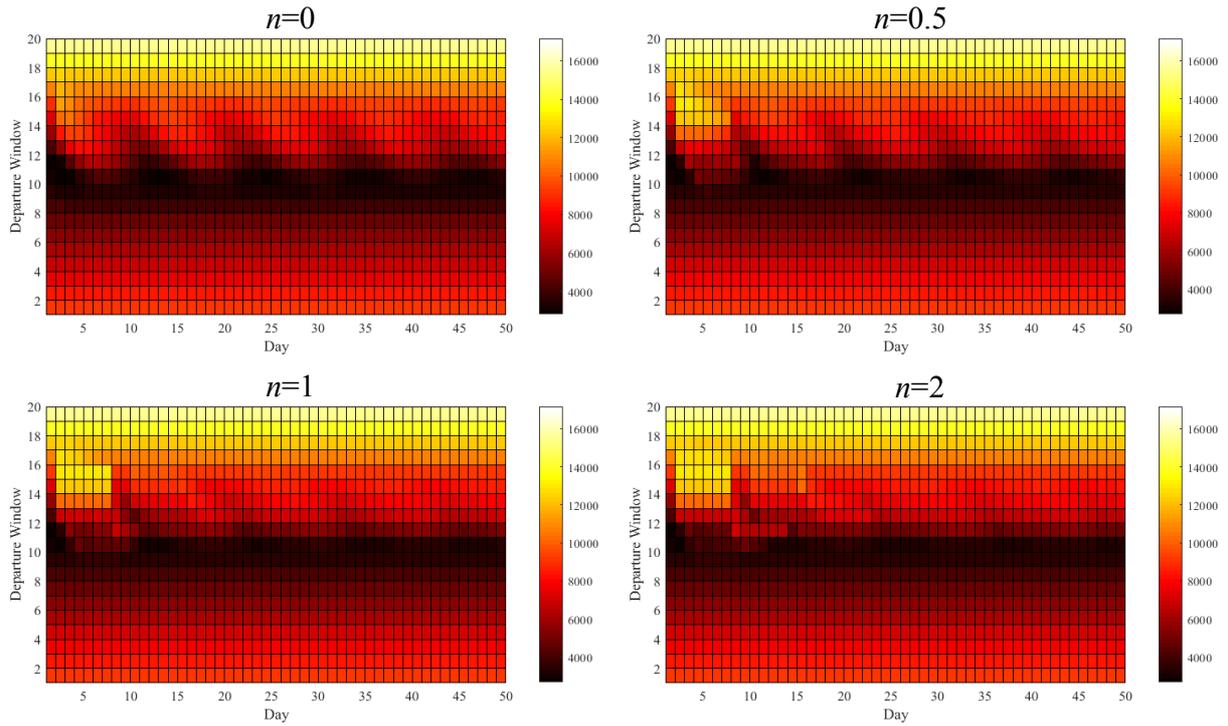

Figure 11. Average perceived travel costs between O-D pair #4 (routes #39-219) for each departure window for a 50-day simulation.

### 5.2. Route and departure time choices under network disruption

In this section, we simulate a local disruption in the Sioux Falls network by reducing 1/3 of the flow capacity of link #68 (see Figure 4) between day 51 and day 100. After the $100^{th}$ day, its capacity is restored. Link 68 is chosen as it carries the highest traffic volume in the DTD simulation reported in Section 5.1.1, and is therefore considered a critical component of the network supply. We use the Base Model II + IS from Section 5.1.1 to illustrate the SRDT choices in the DTD dynamics.

Figure 12 illustrates the change in SRDT choices after the disruption took place. Figure 12(a) shows two departure peaks around $8^{th}$-$9^{th}$ window and $12^{th}$-$13^{th}$ window, respectively before the disruption. After the $51^{st}$ day, there is a clear shift of departure times towards earlier windows in response to the congestion created by the local disruption. Figure 12(b) shows the route choice switches caused by the local disruption. In this figure, paths #3321-3335 and #3336-3346 belong to two different O-D pairs. When disruption occurs, we can clearly observe travelers switching routes and, more interestingly, on certain days they switch back to their original routes, displaying a periodic switching pattern.

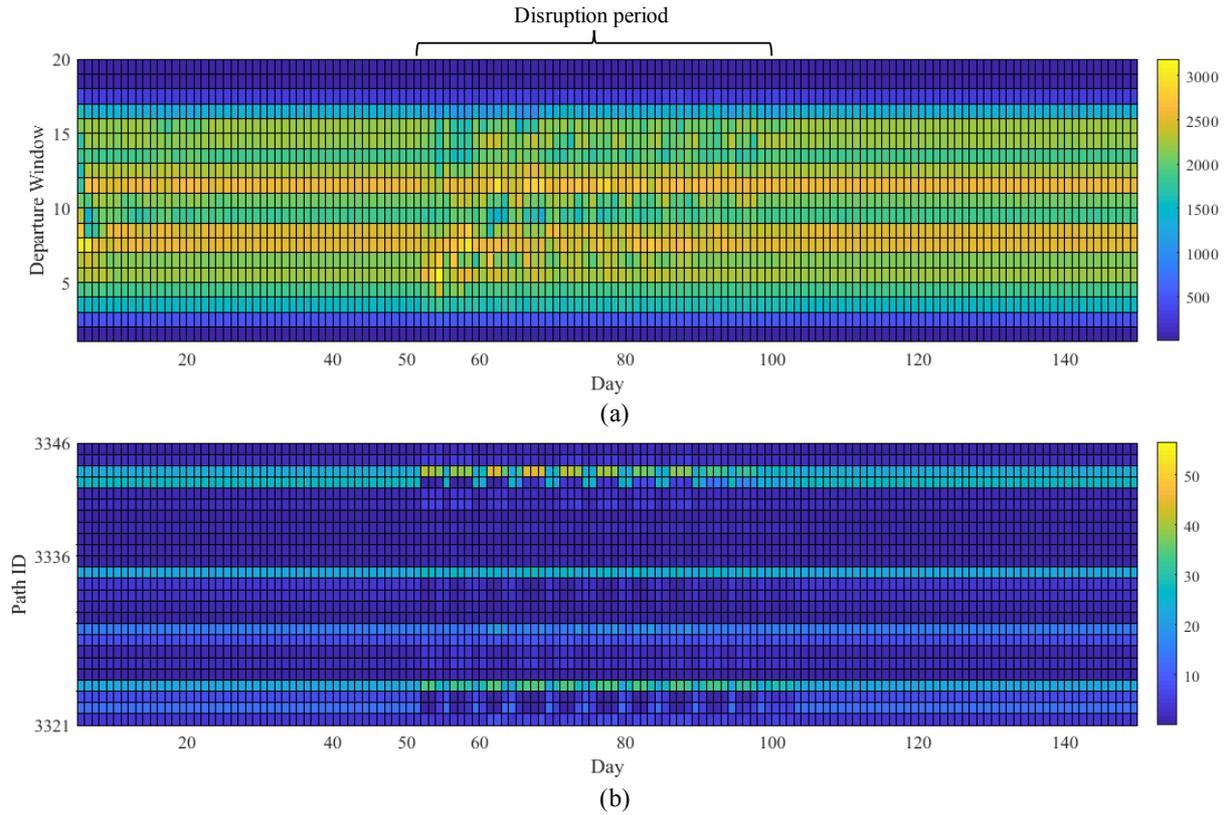

Figure 12. (a): the total departure volumes in each departure window on different days. (b): the total departure volumes along routes 3321-3346 on different days.

We further analyze O-D pairs directly impacted by the disruption. Hereafter, we consider an O-D pair directly impacted by the disruption if the O-D contains at least one route that traverses the disrupted link. Figure 13 shows the daily departure volumes along four routes in one such O-D pair. It can be seen that for routes # 3355 and # 3358, which both traverse link 68, there is a drastic decrease of route volumes at window 6 between day 51 and 100. Those flows are switched to routes #3354 and #3357, which circumvent link 68. This clearly shows the route switching behavior within the impacted O-D pair. Furthermore, some traffic is seen to switch to departure window 5 during the disrupted period for all the four routes, which is captured by the proposed SRDT choice model.

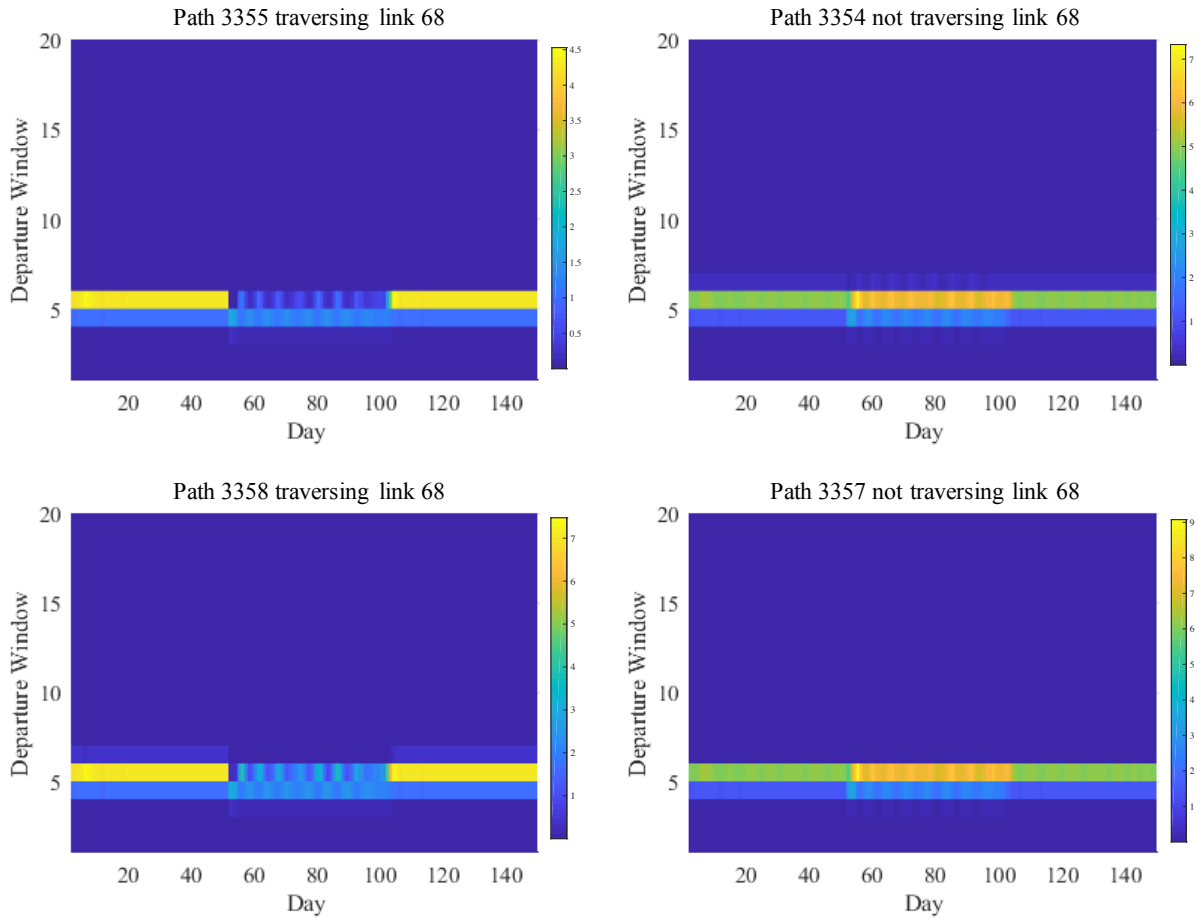

Figure 13. Changes in the SRDT choices after local disruption on day 50.

To further illustrate the departure time shifts, we use Figure 14 to show the cumulative departures of all O-D pairs directly impacted by the disruption, within four days since the disruption first took place. Both Base Model II with and without information sharing (IS) predict higher cumulative departures at any point in time, which suggests a shift towards earlier departures for these O-D pairs. This is likely caused by the spatial propagation of congestion triggered by vehicle spillback, rendering alternative routes within the same departure period unattractive. In addition, such a trend of earlier departure is more pronounced in the absence of complete information (Base Model II + IS). This suggests that incomplete information produces higher uncertainties in travelers' perceived costs, and hence induces earlier departures to accommodate the expected congestion and uncertainties. Such an interaction between travel information uncertainty and departure time choices is not captured by existing models in the literature. This highlights the need to (a) simultaneously model route and departure time choices; and (b) accurately represent congestion propagation with realistic vehicle queuing dynamics.

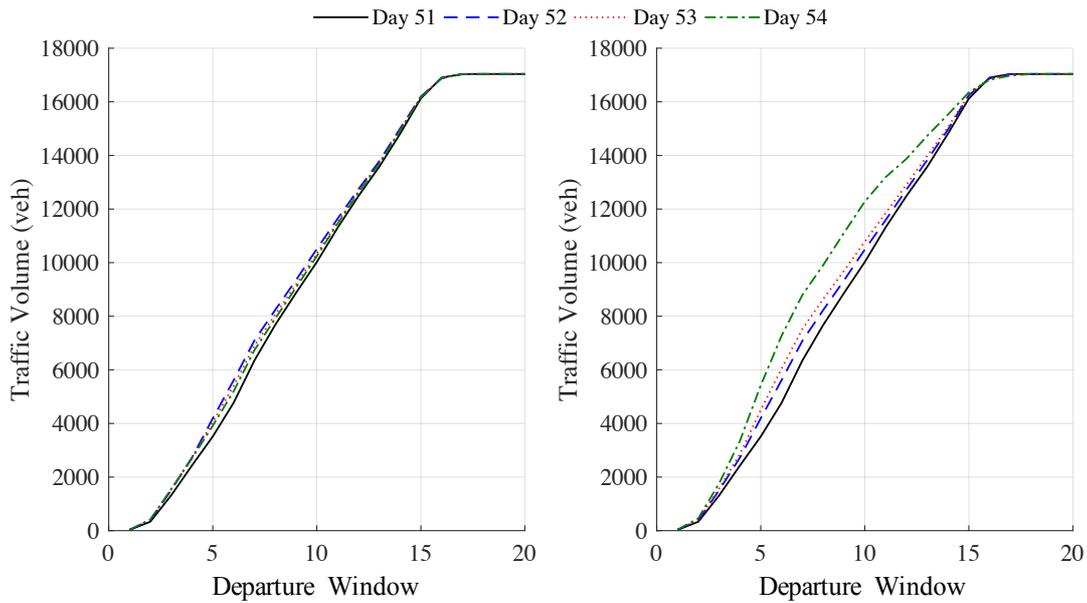

Figure 14. Cumulative departure volume for O-D pairs directly impacted by the local disruption. Left figure: Base Model II, right figure: Base Model II with information sharing.

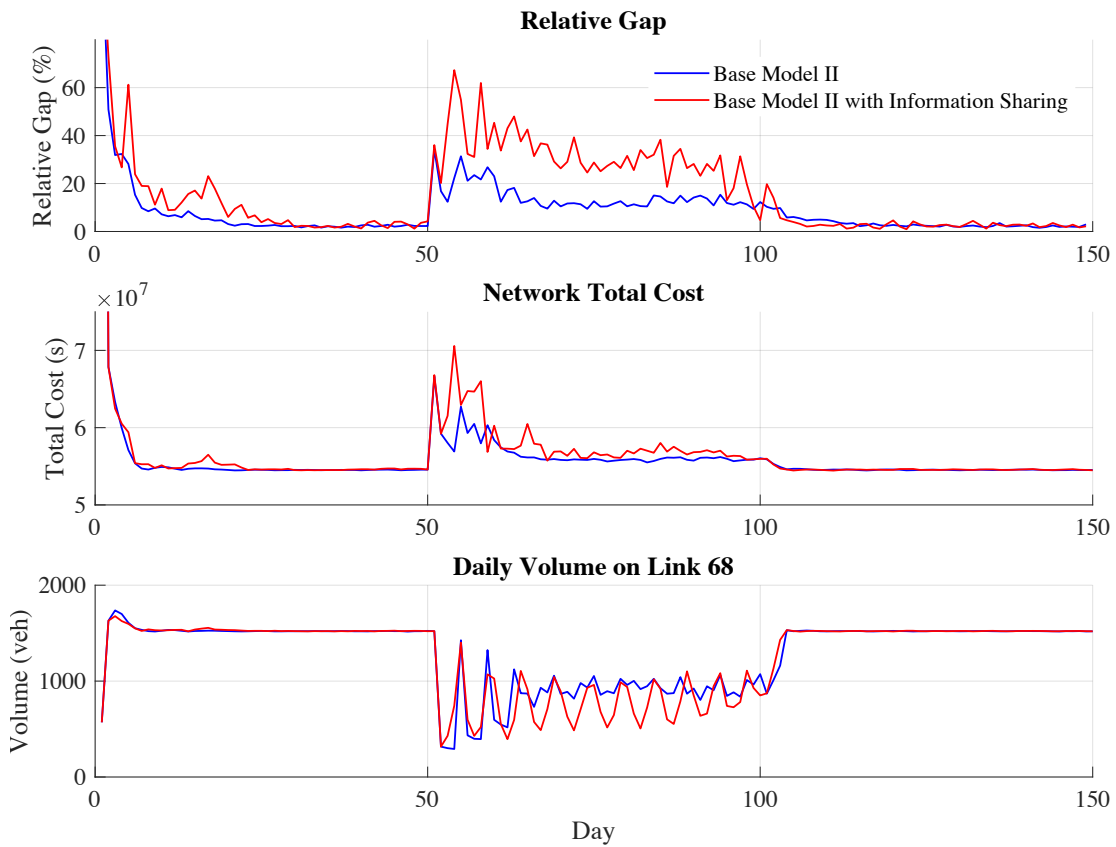

Figure 15. Comparison of Base Model II and Base Model II with information sharing. Top: relative gaps when the network is locally disrupted during $51^{st}$-$100^{th}$ day. Middle: total network cost. Bottom: daily traffic volume.

Finally, for the simulated network disruption, Figure 15 compares Base Model II with and without information sharing, where $n = 2$ for information sharing. As suggested in Figure 8, having access to complete information on every alternative tends to reduce traffic oscillations due to uncertainties, and this remains the case for network disruption, as shown in the top figure of Figure 15: both models predict an increase of the relative gaps when the network is undergoing disruption. Moreover, the traffic does not return to an approximate stationary state for the entirety of the disruption, as the travelers are constant switching routes and departure times, as confirmed by Figure 12.

In terms of the travel costs encountered by all the travelers in the network, both models predict a drastic increase during the disruption period; see the middle figure of Figure 15. It is interesting to observe that having incomplete information also introduces inefficiencies to the entire network. However, the cost gap between the two models diminishes in time and almost vanish right before the 100$^{th}$ day. This means that the lack of complete information leads to a transient state with larger daily traffic variations and higher network-wide cost, and it may take a long time before the system returns to an approximate stationary state.

The bottom picture of Figure 15 shows that the traffic that uses the disrupted link #68 experiences daily oscillations under both models. The magnitude of oscillation predicted by Base Model II is in general lower than that of Base Model II + IS, and tends to be stabilized since around day 70. In contrast, Base Model II + IS is associated with cyclic oscillatory patterns between 600 (veh) and 1000 (veh).

### 5.3. Model variants

In the previous numerical demonstrations, we have tested the model's sensitivity with respect to some of the endogenous parameters such as the scale parameter of the multinomial Logit model, the indifference band, and the information communication parameter $n$ (as in $G(x) = x^n$). In this section we further explore the variability and consistency of the model outputs with respect to some relatively exogenous choices of modeling components, including the route set, the choice of $G(x)$ and the weights described in the departure time choice model (6) and (9).

As we mentioned in the remark at the end of Section 3.1, the set of routes (of size 6180) is generated via the Frank-Wolfe algorithm. To illustrate the model performance with different route sets, we consider three additional route sets with decreasing sizes: 5537, 5270 and 4895. The link disruption experiment in the Sioux Falls network is repeated for these cases (with Base Model II, $\theta = \theta_1 = 0.004, \eta = 400, N = 3$), and the results in terms of network total cost and relative gaps. We can see from Figure 16 that all four cases (including the original case) show qualitatively similar trends in network evolution, suggesting that different route sets lead to similar behavior of the model. It is interesting to note that the overall network cost increases as the number of routes (choice alternatives) reduces, which makes sense as travelers are competing for resources (road capacity) on more stringent choice spaces. The fact that the cases of 5270, 5537 and 6180 routes do not differ significantly suggests that the current choice of 6180 routes is reasonable in representing attractive routes associated with the given travel demands.

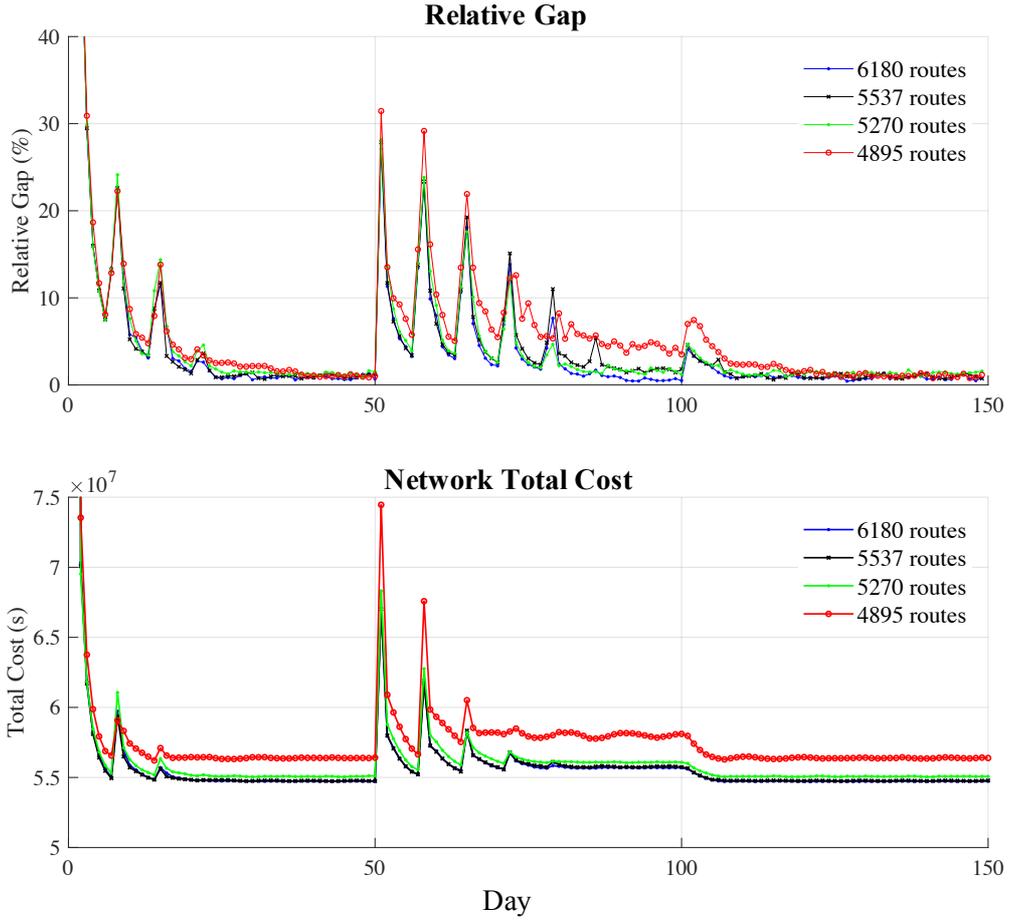

Figure 16. Comparison of different route sets in the Sioux Falls network under link disruption (Base Model II)

Next, we consider two additional forms of $G(x)$ in the information sharing model. The purpose is to demonstrate consistency of the model outputs with regard to different functional forms of $G(x)$, as long as they satisfy the minimum requirement of being monotonic, and $G(0) = 0, G(1) = 1$. In deriving the following functional forms, we use $G(x) = x^2$ as the benchmark.

1. **Piecewise affine**. We propose the following formula:

$$G(x) = \begin{cases} \frac{2}{3}x & x \in [0, 0.75] \\ 2x - 1 & x \in (0.75, 1] \end{cases}$$

2. **Trigonometric functions.** We consider the following formula:

$$G(x) = \tan(1.2x)/\tan(1.2) \quad x \in [0, 1]$$

The two new functions are compared with $G(x) = x^2$ in Figure 17. Next, we run the models (Base Model II with information sharing) and compare the results in Figure 18. We can see that despite the different forms of $G(x)$, the resulting daily trajectories in terms of relative gap and network total costs are quite similar. This verifies the model's consistent performance, which is not significantly affected by the choice of $G(x)$ as long as it satisfies $G(0) = 0, G(1) = 1$ and being monotonically increasing.

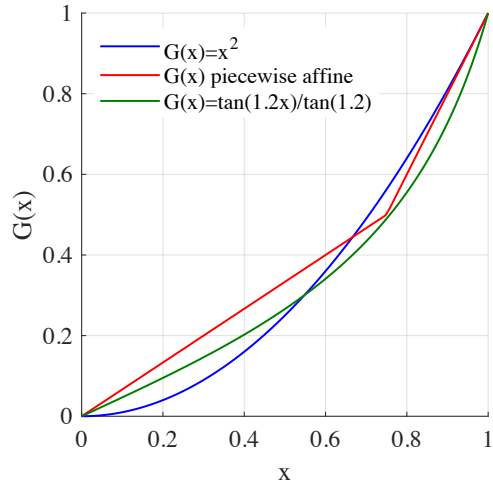

Figure 17. Different forms of $G(x)$.

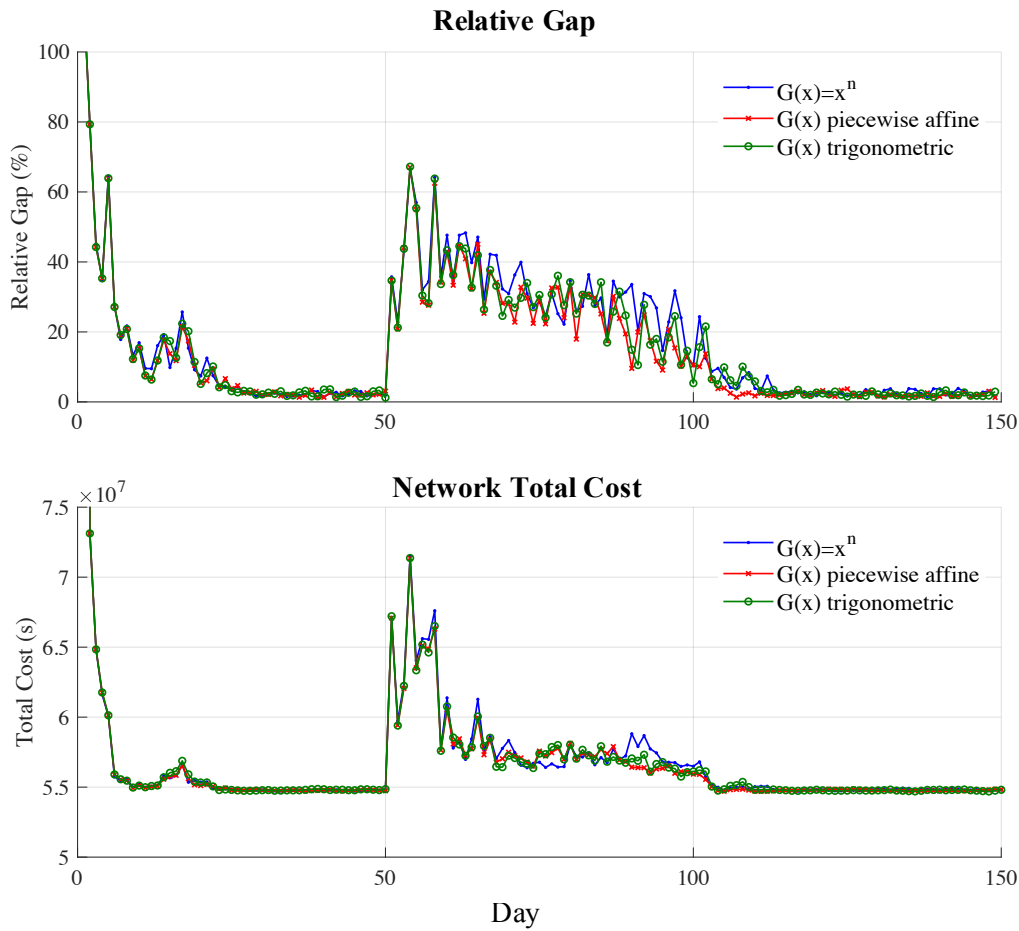

Figure 18. Comparison of different functional forms of $G(x)$ in the Sioux Falls network under link disruption (Base Model II with information sharing)

Finally, we compare the departure time window choice models within Base Model II. Two variants, based respectively on arithmetic mean (6) and harmonic mean (9) of the route costs within the same time window, are proposed. The arithmetic mean considers all the route costs with equal weights, while the harmonic mean places more weights on the more efficient routes. As can be seen in Figure 19, the two cases differ in that

1. The case with harmonic mean leads to reduced total network costs at both equilibrated and disequilibriated conditions. This may be due to travelers' perception of a departure window being influenced by a few attractive routes, instead of all the possible routes. This allows travelers to more effectively explore the spatial-temporal network capacity.
2. The case with harmonic mean results in less oscillations in the network during the disruption period. A possible cause is that their decisions on departure time window are influenced by only a subset of all possible routes.

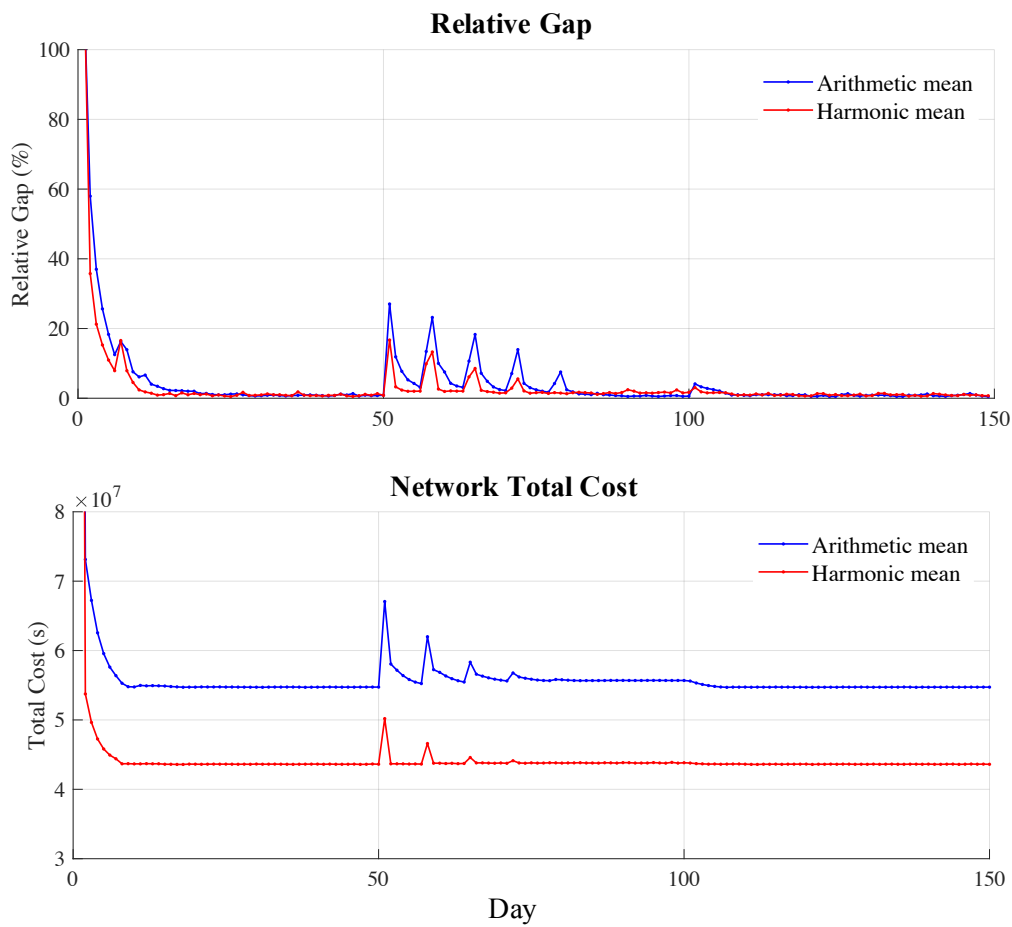

Figure 19. Comparison of Base Model II with different departure time choice models.

## 6. Discussion and Conclusion

This paper presents a doubly dynamic, day-to-day traffic assignment model with realistic user behavior pertaining to imperfect and incomplete information as well as bounded rationality. In particular, we propose two stochastic DTD models where travelers choose departure time and routes based on their adaptive learning and decision making. The first model is based on a multinomial Logit model where each route-and-departure-time pair is treated as an independent alternative. The second model is derived by viewing the departure time and route as sequential choices, which are respectively captured by multinomial Logit and nested Logit models. The nested Logit model for route choices also corrects the IIA assumption.

Building on these models, we further incorporate bounded rationality (BR) and information sharing mechanism into the macroscopic choice modeling framework. The BR assumes that travelers are reluctant to change their previous day's choice unless a gain larger than a threshold (indifference band) is expected. We invoke the random utility theory with indifference band and extend adapt it to the DTD SRDT choice update. On the other hand, travel information availability and reliability are incorporated into the choice model. This is done by assuming that the weight of information on certain alternative reported by a crowd depends on the number (or percentage) of travelers who chose that alternative. A simple yet insightful functional form is provided to parameterize the strength of information sharing.

The within-day dynamics follow the LWR-based dynamic network loading (DNL), which explicitly captures physical queuing and network-wide propagation of congestion. The DNL model is formulated as a system of difference algebraic equations in discrete time, and has been implemented in large-scale networks including the Sioux Falls (528 O-D pairs, 6,180 routes) and Anaheim (1,406 O-D pairs, 30,719 routes) networks. The following conclusions are made from the numerical results:

1) Daily oscillation of traffic flow, measured by the relative gap, depends on the strength of information communication, which is parameterized by $n \in [0,1]$ in the function $G(x) = x^n$ (Section 3.4). In particular, in the Sioux Falls network, the relative gaps increase with $n$ (Figure 8), which means that network traffic undergoes larger daily oscillation when the travelers only have access to information of a few popular choices; such oscillation is minimized when the travelers have complete information of every alternative.

2) The size of the indifference band ($\delta$) has an impact on the daily oscillation. Larger $\delta$ reduces the relative gaps, which conforms with intuition that travelers' reluctance to switch choices has a stabilizing effect on the daily dynamics.

3) Under certain network conditions, the absence of complete information on every single alternative tends to smooth the daily variations of perceived costs, and hence could stabilize daily traffic and reduce congestion cost associated with constant SRDT choice switches (Section 5.1.2). Note that this contradicts the observations made in 1) concerning the Sioux Falls network, which suggests that this type of network behavior is case-dependent.

4) Local disruption (in the form of link capacity reduction in this paper) tends to induce earlier departures for relevant O-D pairs. This is likely caused by the spatial propagation of congestion triggered by vehicle spillback, rendering alternative routes within the same departure window unattractive. This highlights the need to (a) simultaneously model route and departure time choices; and (b) accurately represent congestion propagation with realistic vehicle queuing dynamics.

5) Relating to 4), the amount of shift towards earlier departure windows depends on the information shared among travelers. In particular, larger $n$ (as in $G(x) = x^n$) leads to more shift. This is caused by travelers' conservative behavior when facing uncertainties in the absence of complete and accurate travel information.

6) Relating to 5), the model displays similar behavior when the functional form of $G(x)$ undergo minor perturbations. For example, assuming that $G(x)$ is increasing, convex and satisfies $G(0) = 0, G(1) = 1$ (examples include $G(x) = x^2$, piecewise affine and trigonometric functions), the model results are exhibit qualitatively similar patterns.

7) Under local disruption, the network may undergo a significant transient period before reaching the (approximate) stationary state. This has been corroborated by Figure 15. The transient state may be associated with daily traffic variations and considerable congestion and social cost. Unfortunately, such transient states have been overlooked in many equilibrium-based traffic network design and optimization approaches, such as those based on Stackelberg games and Mathematical Program with Equilibrium Constraints.

8) When making departure time choices in Base Model II, travelers (with complete information on alternative routes in a particular window) may evaluate different routes with equal weighting (Eqn (6)), or give higher weights to the better routes (Eqn (9)). Numerical experiment suggests that the latter leads to less network oscillation and higher efficiency, as travelers' perception of a time window depends only on a few efficiency routes instead of all possible ones.

9) Although the proposed models comprise numerous components and parameters, which makes it difficult to derive theoretical results on their performances and behavior, it is shown through numerical examples and extensive sensitivity analyses that the model outputs remain consistent and stable with respect to different choices of their parameters and components. In addition, the model behavior coincides with intuition. This makes these models desirable to study and predict complex traffic phenomena such as network disequilibria, for which no existing modeling tools are available.

Findings of this work, including model development, network insights and open-access codes, will be of interest to network modelers aiming to predict and quantify the network effect of internal or external changes. This is particularly relevant to emerging and disruptive technologies that impact the system on demand or supply sides, such as adaptive traffic controls, shared mobility, and real-time information dissemination.

## Appendix. The discretized DAE system for dynamic network loading model

The LWR-based DNL procedure is formulated as a system of differential algebraic equations, and we present its discretized version here without going through the detailed derivation. For more extensive discussion and numerical examples, the reader is referred to Han et al. (2019). The following notations are needed to present the DNL model.

- $S$: Set of origins in the network
- $R$: Set of routes employed by all travelers
- $R^o$: Set of routes originating from $o \in S$
- $I^J$: Set of incoming links of a junction $J$
- $O^J$: Set of outgoing links of a junction $J$
- $A^J(s)$: Flow distribution matrix of junction $J$, which is time-dependent
- $f_r(s)$: Route departure rate along $r \in R^w$ at time $s$
- $\text{TT}_r(s)$: Travel time along route $r$ with departure time $s$
- $f_i^{\text{in}}(s)$: Inflow of link $i$
- $f_i^{\text{out}}(s)$: Outflow of link $i$
- $N_i^{\text{up}}(s)$: Link $i$'s cumulative entering count

$N_i^{dn}(s)$: Link $i$'s cumulative exiting count

$D_i(s)$: Demand of link $i$

$S_i(s)$: Supply of link $i$

$\mu_i^r(s)$: Percentage of flow at the entrance of link $i$ associated with route $r$

$q_o(s)$: Point queue at the origin node $o \in S$

$\xi_i(s)$: Entry time of link $i$ corresponding to exit time $s$

$\zeta_i(s)$: Exit time of link $i$ corresponding to entry time $s$

$ds$: Time step size for the dynamic network loading

$L_i, C_i, v_i, u_i, \rho_i^{jam}$: Length, capacity, forward wave speed, backward wave speed, and jam density of link $i$ (assuming triangular fundamental diagram)

We present the DNL procedure in discrete time as follows.

$$D_o(s+1) = \begin{cases} M & \text{if } q_o(s) > 0 \\ \sum_{r \in R^o} f_r(s) & \text{if } q_o(s) = 0 \end{cases} \quad o \in S \tag{28}$$

$$D_i(s+1) = \begin{cases} f_i^{in}(s - L_i/v_i) & \text{if } N_i^{up}(s - L_i/v_i) \leq N_i^{dn}(s) \\ C_i & \text{if } N_i^{up}(s - L_i/v_i) > N_i^{dn}(s) \end{cases} \tag{29}$$

$$S_j(s+1) = \begin{cases} f_j^{out}(s - L_j/u_j) & \text{if } N_j^{up}(s) \geq N_j^{dn}(s - L_j/u_j) + \rho_j^{jam} L_j \\ C_j & \text{if } N_j^{up}(s) < N_j^{dn}(s - L_j/u_j) + \rho_j^{jam} L_j \end{cases} \tag{30}$$

$$N_i^{dn}(s) = N_i^{up}(\xi_i(s)), \quad N_i^{up}(s) = N_i^{dn}(\zeta_i(s)) \tag{31}$$

$$\mu_j^r(s) = \frac{f_i^{out}(s)\mu_i^r(\xi_i(s))}{f_j^{in}(s)} \quad \forall r \text{ s.t. } \{i,j\} \subset r \tag{32}$$

$$A^J(s) = \{\alpha_{ij}(s)\}, \quad \alpha_{ij}(s) = \sum_{r \ni i,j} \mu_i^r(\xi_i(s)) \tag{33}$$

$$\left([f_i^{out}(s)]_{i \in I^J}, [f_j^{in}(s)]_{j \in O^J}\right) = \Theta\left([D_i(s)]_{i \in I^J}, [S_j(s)]_{j \in O^J}; A^J(s)\right) \tag{34}$$

$$q_o(s+1) = q_o(s) + ds \sum_{r \in R^o} f_r(s) - \min\{D_o(s), S_j(s)\} \tag{35}$$

$$N_i^{up}(s+1) = N_i^{up}(s) + ds \cdot f_i^{in}(s), \quad N_i^{dn}(s+1) = N_i^{dn}(s) + ds \cdot f_i^{out}(s) \tag{36}$$

$$TT_r(s) = \zeta_o \circ \zeta_1 \circ \ldots \circ \zeta_K(s), \quad \forall r = \{o, 1, \ldots, K\} \tag{37}$$

Equation (28) defines the demand at the origin $o$, to be utilized later in (35) to determine the queuing dynamics at the origins following a Vickrey type model. (29) and (30) respectively express the link demand and supply using the variational formulation (Han et al., 2016). They are also key to capture vehicle spillback and inter-link congestion propagation. (31) is known as the flow propagation constraint (Friesz et al., 2011), which defines the link entrance and exit time functions. (32) and (33) together determine the link distribution matrix at the junction based on the first-in-first-out principle. Such a matrix serves as an input of the junction dynamic model (34), where the inflows and outflows of incident links are jointly determined by their respective demands and supplies, via the Riemann Solver (Garavello et al., 2016). (36) updates the link cumulative entering/exiting counts by definition. Finally, (37) defines the path travel time based on individual link travel times using the composition $\circ$ of functions, namely $y_1 \circ y_2(s) \doteq y_2(y_1(s))$.

**References**

Avineri, E., Prashker, J. 2004. Violations of expected utility theory in route-choice stated preferences: certainty effect and inflation of small probabilities. Transportation Research Record: Journal of the Transportation Research Board 1894 (1), 222-229.

Balijepalli, N.C., Watling, D.P., Liu, R. 2007. Doubly Dynamic Traffic assignment: Simulation Modeling Framework and experimental results. Transportation Research Board of the National Academies, Washington, D.C., 39-48.

Ben-Akiva, M. and Bierlaire, M., 2003. Discrete choice models with applications to departure time and route choice. In *Handbook of transportation science* (pp. 7-37). Springer, Boston, MA.

Bie, J., Lo, H. 2010. Stability and attraction domains of traffic equilibria in a day-to-day dynamical system formulation. Transportation Research Part B 44 (1), 90-107.

Bierlaire, M. and Frejinger, E., 2005. Route choice models with subpath components. In *Swiss Transportation Research Conference* (No. TRANSP-OR-CONF-2006-032).

Bifulco, G.N., Cantarella, G.E., Simonelli, F. and Velonà, P., 2016. Advanced traveller information systems under recurrent traffic conditions: Network equilibrium and stability. Transportation Research Part B, 92, 73-87.

Bliemer, M.C.J., Raadsen, M.P.H., Brederode, L.J.N., Bell, M.G.H., Wismans, L.J.J., Smith, M.J. 2017. Genetics of traffic assignment models for strategic transport planning. Transport reviews 37(1), 56-78.

Boyce, D., Lee, D.H., Ran, B. 2001. Analytical models of the dynamic traffic assignment problem. Networks and Spatial Economics, 1, 377-390.

Cantarella, G.E., Astarita, V. 1999. A Doubly Dynamic Traffic Assignment Model for Planning Applications. Conference: 14th International Symposium on transportation and Traffic Theory, At: Jerusalem.

Cantarella, G.E, Cascetta, E. 1995. Dynamic processes and equilibrium in transportation networks: towards a unifying theory. Transportation Science 29 (4), 305-329.

Cantarella, G.E. and Watling, D.P., 2016. A general stochastic process for day-to-day dynamic traffic assignment: formulation, asymptotic behaviour, and stability analysis. Transportation Research Part B, 92, 3-21.

Cascetta, E. 1989. A stochastic process approach to the analysis of temporal dynamics in transportation networks, Transportation Research Part B 23, 1-17.

Cascetta, E., Cantarella, G.E., 1991. A day-to-day and within-day dynamic stochastic assignment model. Transportation Research Part A 25, 277-291.

Cascetta, E., Cantarella, G.E. 1993. Modelling dynamics in transportation networks, Journal of Simulation Practice and Theory- Elsevier.

Daganzo, C.F., 1994. The cell transmission model: A simple dynamic representation of highway traffic. Trans. Res. B 28B (4), 269-287.

Daganzo, C.F., 1995. The cell transmission model, part II: network traffic. Transportation Research Part B 29B, 79-93.

Di, X., Liu, H.X., Ban, X., Yu, J.W., 2015. Stability of a boundedly rational day-to-day dynamic. Networks and Spatial Economics 15(3), 537-557.

Friesz, T.L., Bernstein, D., Smith, T., Tobin, R., Wie, B., 1993. A variational inequality formulation of the dynamic network user equilibrium problem. Operations Research 41 (1), 80-91.

Friesz, T.L., Bernstein, D., Mehta, N.J., Tobin, R.L. and Ganjalizadeh, S., 1994. Day-to-day dynamic network disequilibria and idealized traveler information systems. Operations Research, 42(6), 1120-1136.

Friesz, T.L., Cho, H., Mehta, N., Tobin, R., 1992. Simulated annealing methods for network design problems with variational inequality constraints. Transportation Science 26 (1), 18-26.

Friesz, T.L., Kim, T., Kwon, C. and Rigdon, M.A., 2011. Approximate network loading and dual-time-scale dynamic user equilibrium. Transportation Research Part B, 45(1), 176-207.

Friesz, T.L., Tobin, R., Shah, S., Mehta, N., Anandalingam, G., 1993. The multiobjective equilibrium network design problem revisited: a simulated annealing approach. European Journal of Operations Research 65, 44-57.

Garavello, M., Han, K., Piccoli, B., 2016. Models for Vehicular Traffic on Networks. American Institute of Mathematical Sciences.

Ge, Y.E., Zhou, X. 2012. An alternative definition of dynamic user equilibrium on signalized road networks. Journal of Advanced Transportation 46, 236-253.

Gifford, J.L., Checherita, C. 2007. Bounded rationality and transportation behavior: lessons for public policy. In: Transportation Research Board 86th Annual Meeting, No. 07-2451.

Guo, R.Y. and Szeto, W.Y., 2018. Day-to-day modal choice with a Pareto improvement or zero-sum revenue scheme. Transportation Research Part B, 110(C), 1-25.

Guo, R.Y., Yang, H., Huang, H.J. and Li, X., 2017. Day-to-day departure time choice under bounded rationality in the bottleneck model. Transportation research procedia, 23, 551-570.

Guo, R.Y., Yang, H., Huang, H.J., Tan, Z., 2015. Link-based day-to- day network traffic dynamics and equilibria. Transportation Research Part B 71, 248-260

Guo, X. and Liu, H.X., 2011. Bounded rationality and irreversible network change. Transportation Research Part B, 45(10), 1606-1618.

Han, K., Eve, G. and Friesz, T., 2019. Computing dynamic user equilibria on large-scale networks with software implementation. Networks and Spatial Economics, 19(3), 869-902.

Han, K., Piccoli, B. and Szeto, W.Y., 2016. Continuous-time link-based kinematic wave model: formulation, solution existence, and well-posedness. Transportmetrica B: Transport Dynamics, 4(3), 187-222.

Han, K., Szeto, W.Y. and Friesz, T.L., 2015. Formulation, existence, and computation of boundedly rational dynamic user equilibrium with fixed or endogenous user tolerance. Transportation Research Part B, 79, 16-49.

Han, Q., Timmermans, H. 2006. Interactive learning in transportation networks with uncertainty, bounded rationality, and strategic choice behavior: quantal response model. Transportation Research Record: Journal of the Transportation Research Board 1964 (1), 27-34.

Hazelton, M., Watling, D. 2004. Computation of equilibrium distributions of Markov traffic assignment models. Transportation Science, 38 (3), 331-342.

He, X., Guo, X., Liu, H. 2010. A link-based day-to-day traffic dynamic model. Transportation Research Part B 44 (4), 597-608.

He, X., Liu, H.X. 2012. Modeling the day-to-day traffic evolution process after an unexpected network disruption. Transportation Research Part B 46 (1), 50-71.

Horowitz, J.L. 1984. The stability of stochastic equilibrium in a two-link transportation network. Transportation Research Part B 18, 13-28.

Hu, T., Mahmassani, H. 1997. Day-to-day evolution of network flows under real-time information and reactive signal control. Transportation Research Part C 5 (1), 51-69.

Iryo, T. 2016. Day-to-day dynamical model incorporating an explicit description of individuals' information collection behavior. Transportation Research Part B 92, 88-103.

Krishnan, K.S., 1977. Incorporating thresholds of indifference in probabilistic choice model. Management Science 23(11), 1224-1233.

LeVeque, R.J., Numerical Methods for Conservation Laws. 1992. Birkhauser Basel.

Li, X., Liu, W., Yang, H., 2018. Traffic dynamics in a bi-model transportation network with information provision and adaptive transit services. Transportation Research Part C 91, 77-98.

Lighthill, M., Whitham, G., 1955. On kinematic waves. II. A theory of traffic flow on long crowded roads. Proceedings of the Royal Society of London: Series A 229, 317- 345.

Long, J., Chen, J., Szeto, WY., Shi, Q. 2018. Link-based system optimum dynamic traffic assignment problems with environmental objectives. Transportation Research Part D 60, 56-75.

Lou, Y., Yin, Y., Lawphongpanich, S. 2010. Robust congestion pricing under bounded rational user equilibrium. Transportation Research Part B 44 (1), 15-28.

Liu, W., Li, X., Zhang, F., Yang, H., 2017. Interactive travel choices and traffic forecast in a doubly dynamical system with user inertia and information provision. Transportation Research Part C 85, 711-731.

Mahmassani, H., Chang, G., 1987. On boundedly rational user equilibrium in transportation systems. Transportation Science 21 (2), 89-99.

Mahmassani, H., Jayakrishnan, R. 1991. System performance and user response under real-time information in a congested traffic corridor. Transportation Research Part A 25 (5), 293-307.

Mahmassani, H., Liu, Y. 1999. Dynamics of commuting decision behavior under advanced traveller information systems. Transportation Research Part C 7(2-3), 91-107.

Mahmassani, H., Zhou, X., Lu, C. 2005. Toll pricing and heterogeneous users: approximation algorithms for finding bicriterion time-dependent efficient paths in large-scale traffic networks. Transportation Research Record: Journal of the Transportation Research Board 1923 (1), 28-36.

Nagurney, A. and Zhang, D., 1997. Projected dynamical systems in the formulation, stability analysis, and computation of fixed-demand traffic network equilibria. Transportation Science, 31(2), 147-158.

Nie, X., Zhang, H.M., 2005. A comparative study of some macroscopic link models used in dynamic traffic assignment. Networks and Spatial Economics 5, 89-115.

Osorio, C., G. Flotterod, and M. Bierlaire 2011. Dynamic network loading: A stochastic differentiable model that derives link state distributions. Transportation Research Part B 45 (9): 1410-1423.

Ouyang, Y. 2007. Pavement resurfacing planning for highway networks: parametric policy iteration approach. Journal of Infrastructure Systems 13(1), 65-71.

Parry, K., Hazelton, M.L. 2013. Bayesian inference for day-to-day dynamic traffic models. Transportation Research Part B 50, 104-115.

Peeta, S., Ziliaskopoulos, A.K. 2001. Foundations of dynamic traffic assignment: The past, the present and the future. Networks and Spatial Economics 1, 233-265.

Rambha T. and Boyles S. D. 2016. Dynamic pricing in discrete time stochastic day-to-day route choice models. Transportation Research Part B, 92, 104-118.

Ran, B., Boyce, D., 1996. A link-based variational inequality formulation of ideal dynamic user optimal route choice problem. Transportation Research Part C 4 (1), 1-12.

Richards, P.I., 1956. Shockwaves on the highway. Operations Research 4, 42-51.

Ridwan, M. 2004. Fuzzy preference based traffic assignment problem. Transportation Research Part C 12 (3), 209-233.

Shang, W., Han, K., Ochieng, W. and Angeloudis, P., 2017. Agent-based day-to-day traffic network model with information percolation. Transportmetrica A: Transport Science, 13(1), 38-66.

Simon, H.A. 1957. A behavioral model of rational choice. In: Models of Man, Social and Mathematical Essays on Rational Human Behavior in a Social Setting. Wiley, New York.

Small, K.A., 1982. The scheduling of consumer activities: work trips. American Economic Review 72, 467-479.

Smith, M.J., 1984. The stability of a dynamic model of traffic assignment - an application of a method of Lyapunov. Transportation Science, 18(3), 245-252.

Smith, M.J., Mounce, R. 2011. A splitting rate model of traffic rerouting and traffic control. Transportation Research Part B 45 (9), 1389-1409.

Szeto, W.Y., Jiang, Y. 2011. A Cell-Based Model for Multi-class Doubly Stochastic Dynamic Traffic Assignment. Computer-Aided Civil and Infrastructure Engineering 26, 595–611.

Szeto WY, Lo HK (2004) A cell-based simultaneous route and departure time choice model with elastic demand. Transportation Research Part B 38(7), 593-612

Szeto, W.Y., Lo, H.K. 2005. Dynamic Traffic assignment: review and future research directions. Journal of Transportation Systems Engineering and Information Technology 5(5), 85-100.

Szeto, W.Y., Lo, H.K. 2006. Dynamic traffic assignment: Properties and extensions. Transportmetrica 2 (1), 31-52.

Wang, J., He, X. and Peeta, S., 2016. Sensitivity analysis based approximation models for day-to-day link flow evolution process. Transportation Research Part B, 92, 35-53.

Wardrop, J. 1952. Some theoretical aspects of road traffic research. In ICE Proceedings: Part II, Engineering Divisions 1, 325-362.

Watling, D. P. 1996. Asymmetric problems and stochastic process models of traffic assignment. Transportation Research Part B 30 (5), 339-357.

Watling, D. P. 1999. Stability of the stochastic equilibrium assignment problem: a dynamical systems approach. Transportation Research Part B 33 (4), 281-312.

Watling, D. P., Cantarella, G.E. 2013. Model representation & decision-making in an ever-changing world: the role of stochastic process models of transportation systems. Networks and Spatial Economics in press.

Watling, D. P., Cantarella, G.E. 2015. Model representation & decision-making in an ever-changing world: the role of stochastic process models of transportation systems. Networks and Spatial Economics 15 (3), 843-882.

Watling, D.P, Hazelton, M.L. 2003. The dynamics and equilibria of day-to-day assignment models.

Networks and Spatial Economics 3 (3), 349-370.

Watling, D.P. and Hazelton, M.L., 2018. Asymptotic approximations of transient behaviour for day-to-day traffic models. Transportation Research Part B, 118, 90-105.

Watling, D.P., Rasmussen, T.K., Proto, C.G., Nielsen, O.A. 2018. Stochastic user equilibrium with a bounded choice model. Transportation Research Part B 114, 245-280.

Wei, F., Jia, N., Ma, S., 2016. Day-to-day traffic dynamics considering social interaction: From individual route choice behaviour to a network flow model. Transportation Research Part B, 94, 335-354.

Xiao, Y., Lo, H. K. 2016. Day-to-day departure time modeling under social network influence. Transportation Research Part B 92, 54-72.

Xu, H., Yang, H., Zhou, J. and Yin, Y., 2017. A route choice model with context-dependent value of time. Transportation Science, 51(2), 536-548.

Yang, F., Zhang, D. 2009. Day-to-day stationary link flow pattern. Transportation Research Part B 43 (1), 119-126.

Ye, H., Yang, H., 2017. Rational behavior adjustment process with boundedly rational user equilibrium. Transportation Science, 51(3), 968-980.

Yperman, I., Logghe, S., Immers, L. (2005). The link transmission model: An efficient implementation of the kinematic wave theory in traffic networks, Advanced OR and AI Methods in Transportation. Proceedings of the 10th EWGT meeting and 16th Mini-EURO conference, Poznan, Poland, 122-127, Publishing House of Poznan University of Technology.

Zhang, C., Liu, T., Huang, H., Chen, J., 2018. A cumulative prospect theory approach to commuters' day-to-day route-choice modeling with friends' travel information. Transportation Research Part C 86, 527-548.

Zhou, B., Xu, M., Meng, Q., Huang, Z., 2017. A day-to-day route flow evolution process towards the mixed equilibria. Transportation Research Part C 82, 210-228.

Zhu, S., Levinson, D.M., 2012. Disruptions to transportation networks: a review. In Network reliability in practice (pp. 5-20). Springer, New York, NY.